\documentclass{aa}
\usepackage[varg]{txfonts}
\usepackage{graphicx}
\usepackage{epstopdf}
\usepackage{pdflscape}
\usepackage{booktabs}
\usepackage{txfonts}
\usepackage{gensymb}
\usepackage{color}
\usepackage{url}
\usepackage{multirow}
\usepackage{amsmath}
\usepackage{amstext}
\usepackage{natbib}
\usepackage{longtable}
\usepackage{float}
\usepackage[export]{adjustbox}

\newcommand{\ud}{\mathrm{d}}

\begin{document}

\title{VLT spectropolarimetry of comet 67P: Dust environment around the end of its intense Southern summer}

\author{Yuna G. Kwon\inst{\ref{inst1}\thanks{Alexander von Humboldt Postdoctoral Fellow}}~\and Stefano Bagnulo\inst{\ref{inst2}} \and Johannes Markkanen\inst{\ref{inst1},\ref{inst3}} \and Jessica Agarwal\inst{\ref{inst1},\ref{inst3}} \and Kolokolova Ludmilla\inst{\ref{inst4}}  \and Anny-Chantal Levasseur-Regourd\inst{\ref{inst5}} \and Colin Snodgrass\inst{\ref{inst6}} \and Gian P. Tozzi\inst{\ref{inst7}}
 }

\institute{Institut f{\" u}r Geophysik und Extraterrestrische Physik, Technische Universit{\" a}t Braunschweig,  Mendelssohnstr. 3, 38106 Braunschweig, Germany\\
\email{y.kwon@tu-braunschweig.de}\label{inst1}
\and Armagh Observatory, College Hill, Armagh BT61 9DG, Northern Ireland, UK\label{inst2}
\and  Max Planck Institute for Solar System Research, Justus-von-Liebig-Weg 3, 37077 G{\"o}ttingen, Germany\label{inst3}
\and Department of Astronomy, University of Maryland, College Park, MD 20742, USA\label{inst4}
\and LATMOS, Sorbonne Universit{\' e}, CNRS, UVSQ, Campus Pierre et Marie Curie, 4 place Jussieu, 75005 Paris, France\label{inst5}
\and Institute for Astronomy, University of Edinburgh, Royal Observatory, Edinburgh EH9 3HJ, UK\label{inst6}
\and INAF -- Osservatorio Astrofisico di Arcetri, 50125, Firenze, Italy\label{inst7}
}

\date{Received August 26, 2021 / Accepted ---}

\abstract {A cornucopia of Rosetta and ground-based observational data sheds light on the evolution of the characteristics of dust particles from comet 67P/Churyumov-Gerasimenko (hereafter 67P) with seasons, implying the different dust environments in the source regions on the surface of the comet. }
{We aim to constrain the properties of the dust particles of 67P and therefrom diagnose the dust environment of its coma and near-surface layer at around the end of the Southern summer of the comet. }
{We performed spectropolarimetric observations for 67P dust over 4,000--9,000 \AA\ using the ESO/Very Large Telescope in January--March 2016 (phase angle ranging $\sim$26\degree--5\degree). We examined the optical behaviours of the dust, which, together with Rosetta colour data, were used to search for dust evolution with cometocentric distance. Modelling was also conducted to identify the dust attributes compatible with the results.}
{The spectral dependence of the polarisation degree of 67P dust is flatter than found in other dynamical groups of comets in similar observing geometry. The depth of its negative polarisation branch appears to be a bit shallower than in long-period comets and might be getting shallower as 67P repeats its apparitions. Its dust colour shows a change in slope around 5,500 \AA, (17.3 $\pm$ 1.4) and (10.9 $\pm$ 0.6) \% (1,000 \AA)$^{\rm -1}$ for shortward and longward of the wavelength, respectively, which are slightly redder but broadly consistent with the average of Jupiter-Family comets.}
{Observations of 67P dust in this study can be attributed to dust agglomerates of $\sim$100 $\mu$m in size detected by Rosetta in early 2016. A porosity of 60 \% shows the best match with our polarimetric results,  yielding a dust density of $\sim$770 kg m$^{\rm -3}$. Compilation of Rosetta and our data indicates the dust's reddening with increasing nucleus distance, which may be driven by water-ice sublimation as the dust moves out of the nucleus. We estimate the possible volume fraction of water ice in the initially ejected dust as $\sim$6 \% (i.e. the refractory-to-ice volume ratio of $\sim$14). }


\titlerunning{Spectropolarimetric study of comet 67P in its 2016 outbound phase}

\authorrunning{Y. G. Kwon et al.}

\maketitle

\section{Introduction \label{sec:intro}}

The two-year rendezvous with comet 67P/Churyumov-Gerasimenko (hereafter 67P) by the ESA/Rosetta spacecraft from 3.60 au inbound on 2014 August 6 to 3.83 au outbound on 2016 September 30 provides an unprecedented database of cometary dust. In particular, the onboard Micro-Imaging Dust Analysis System (MIDAS) unravels a hierarchical nature of the cometary dust: $\gtrsim$100-micrometre dust agglomerates consist of $\sim$tens of micrometre dust aggregates, which are in turn comprised of solid sub-micrometre dust grains (monomers) of similar size as interstellar dust, highlighting the importance of dust evolution indigenous to our planetary system \citep{Bentley2016,Mannel2016,Mannel2019}. The aggregates or agglomerates can be classified by the mechanical strength pertinent to their porosity \citep{Fulle2016,Guttler2019}. The prevailing group of dust particles in the coma varies in the porosity (the volume of voids in a particle) with the comet's season \citep{Longobardo2020}, indicating that the source regions activated in different orbital positions have different properties. Observations showing such seasonal effects thus emphasise the importance of constraining the properties of the ejected dust particles as diagnostics for the dust environment across the nucleus surface (\citealt{Marschall2020} and references therein).

The coordinated ground-based campaign of 67P had supplemented the Rosetta observations throughout the mission phase by monitoring large-scale ($\gtrsim$10$^{\rm 3}$ km in cometocentric distance) comet activity in various observing modes, such as photometry, spectroscopy, and polarimetry, at different spectral domains (see Table 1 in \citealt{Snodgrass2017} for the summary). The campaign aimed to understand better and ultimately forge a comprehensive reference of cometary activity and dust constituents by linking in situ and ground-based observations. As a consequence, the following three aspects of 67P were identified in common by the observations. Firstly, strong seasonal effects of the comet induce discernible asymmetry in macroscopic dust coma morphology and gas production rates between the pre- and post-perihelion phases, with the overall activity peaked at $\lesssim$1 month in its outbound orbit  \citep{Boehnhardt2016,Snodgrass2016,Hadamcik2016,Opitom2017,Ivanova2017,Knight2017}. Secondly, changes in radial profile and polarisation degree of 67P dust on either side of the perihelion indicate that its pre-perihelion dust coma characteristics could be attributed to the presence of large-sized (millimetre-to-centimetre) particles, whereas particles dominating the coma around and just after perihelion would be smaller and/or fluffier dust \citep{Hadamcik2016,Rosenbush2017}. Thirdly, all the above-mentioned traits seem to have remained constant over the last three apparitions \citep{Hadamcik2016,Opitom2017,Knight2017}.

Despite the cogent evidence in the evolution of the global activity pattern of 67P and its dust particles throughout the perihelion passage, the characteristics of the dust, particularly its porosity, are still open to be discussed. A constraint on the dust porosity helps to better understand the mechanisms on the comet surface that shape surface morphology \citep{Longobardo2020,Marschall2020} and macroscopic coma activity \citep{Kwon2019}. However, limitations in the observing geometry of the Rosetta spacecraft that confined it to mostly terminator orbits around the comet, together with limitation in the sensitivity of the onboard dust analysers, put little constraints on the property of interest \citep{Thomas2019}. For ground-based observations, the fast brightness decrease of 67P on its way outward from perihelion and a coarse spatial resolution of the telescopes used (typically $\gtrsim$a few thousand kilometres) made it challenging to obtain high signal-to-noise (S/N) inner coma data \citep{Knight2017}. A study on the nature of the cometary dust under such circumstances thus necessitates adopting bigger diameter telescopes with an efficient observing strategy for the dust study.

At this juncture, polarimetry provides a valuable additional dimension of interpretation on the light scattered by cometary dust \citep{Bohren1983}. Polarisation as an intensity ratio reflects the properties of the scattering dust particles, differing from the intensity in widely used photometry or spectroscopy that is mainly sensitive to the number density of the dust particles within an aperture or a slit of a certain width, which alone tends to provide a degenerate solution on the dust properties. Imaging polarimetry observations of the dust of various comets at different phase angles well showcase the usefulness of polarimetry, showing distinct dust coma structures characterised by different types of dust particles on the polarisation map, which were invisible on the intensity map \citep{Renard1996,Hadamcik2003a}. Spectropolarimetry further provides polarimetric and spectroscopic information simultaneously; thus allows us to obtain the information of the polarisation degree, the position angle of polarisation, their variations about wavelength, and dust colour, free of gas contamination (e.g. Fig. 2 in \citealt{Kwon2018}). Not many studies have yet been made in this observing mode because of the small number of available facilities and small limiting magnitudes \citep{Kiselev2015}.

Here, we present new spectropolarimetric observations of comet 67P from the European Southern Observatory's Very Large Telescope (VLT). The VLT's 8-m diameter and dual-beam polarimetry optics allow us to conduct spectropolarimetric observations of 67P's faint coma ($V$-band apparent magnitude of $\sim$16.34 mag\footnote{https://ssd.jpl.nasa.gov/horizons.cgi\#results}) around the end of its Southern summer, with negligible atmospheric effects. We performed observations at three epochs in about a monthly cadence in early 2016. Combining the results of the Rosetta dust studies with the modelling of our observations, we estimate the porosity of the dust particles of 67P and discuss the possible amount of embedded water ice and the refractory-to-ice ratio of the dust. Sect. \ref{sec:obsdata} describes the observational methods and data analyses, and in Sect. \ref{sec:res}, we present the results, which will be discussed in Sect. \ref{sec:discuss}. Finally, in Sect. \ref{sec:sum}, we present a summary.\\

\section{Observations and data analysis \label{sec:obsdata}}

In this section, we describe the observational circumstances and data analyses. The journal of our observational geometry and instrument settings is summarised in Table \ref{t1}.

\begin{table*}[!t]
\centering
\caption{Observational geometry and instrument settings}
\vskip-1ex
\begin{tabular}{c|c|c|c|c|c|ccccc}
\toprule
Median UT & Telescope/ & \multirow{2}{*}{Mode} & \multirow{2}{*}{$N$} & {Exptime} & \multirow{2}{*}{$X$} & $m_{\rm V}$ & $r_{\rm H}$ & $\Delta$ & $\alpha$ & $\nu$\\
2016+ & Instrument & & & (sec) &  & (mag) & (au) & (au) & (\degree) & (\degree)\\
\midrule
\midrule
\multirow{2}{*}{Jan 12 07:00} & \multirow{6}{*}{VLT/FORS2} & \multirow{6}{*}{PMOS/300V} & \multirow{6}{*}{2} & \multirow{2}{*}{2,240} & 1.57 & \multirow{2}{*}{15.951} & \multirow{2}{*}{2.107} & \multirow{2}{*}{1.574} & \multirow{2}{*}{26.24} & \multirow{2}{*}{92.84} \\
 & & & & & (1.44--1.70) & & & & & \\
\multirow{2}{*}{Feb 04 05:41} & & & & \multirow{2}{*}{2,400} & 1.52 & \multirow{2}{*}{16.267} & \multirow{2}{*}{2.283} & \multirow{2}{*}{1.496} & \multirow{2}{*}{18.47} & \multirow{2}{*}{99.56} \\
 & & & & & (1.41--1.63) & & & & & \\
\multirow{2}{*}{Mar 04 08:42} & & & & \multirow{2}{*}{2,400} & 1.74 & \multirow{2}{*}{16.800} & \multirow{2}{*}{2.502} & \multirow{2}{*}{1.528} & \multirow{2}{*}{5.49} & \multirow{2}{*}{106.73} \\
 & & & & & (1.61--1.97) & & & & & \\
\bottomrule
\end{tabular}
\tablefoot{Top headers: Mode, instrumental settings of spectropolarimetry observation (PMOS) with the grism filter (300V); $N$, number of a pair of one set of exposures; Exptime, total exposure time in seconds; $X$, mean airmass with the range in airmass in the bracket; $m_{\rm V}$, apparent total (nucleus + coma) $V$-band magnitude provided by the NASA/JPL Horizons system; $r_{\rm H}$ and $\Delta$, median heliocentric and geocentric distances in au, respectively; $\alpha$, median phase angle (angle of Sun--comet--observer) in degrees; and $\nu$, median true anomaly in degrees.} 
\label{t1}
\vskip-1ex
\end{table*}

\subsection{Observations \label{sec:obs}}

A three-epoch optical spectropolarimetric observation of comet 67P was conducted from UT 2016 January 12 to March 04 in approximately monthly cadence using a polarimeter attached to the 8.0-m diameter VLT at the Paranal Observatory (70\degr24$\arcmin$10\farcs1W, 24\degr37$\arcmin$31\farcs5S, 2635 m) in Chile.  FORS2 is a multi-purpose optical instrument mounted on the Cassegrain focus of the UT1 telescope at VLT. A dual-beam polarimeter\footnote{http://www.eso.org/sci/facilities/paranal/instruments/fors/doc/VLT-MAN-ESO-13100-1543\_P01.pdf}, consisting of a rotatable half-wave plate (HWP) and a Wollaston prism (WP), is located upstream of two 2k $\times$ 4k CCD detectors with the plate scale of 0\farcs125 pixel$^{\rm -1}$ (with the pixel size of 15 $\times$ 15 $\mu$m$^{\rm 2}$) and the field of view (FoV) of 6\farcm8 $\times$ 6\farcm8. In spectropolarimetry (PMOS) observing mode with the standard resolution (SR) collimator, we used the 2 $\times$ 2 binning for fainter targets such as 67P around this term of observations, yielding the pixel scale of 0\farcs25 pixel$^{\rm -1}$. 

The rotatable HWP introduces a phase-shift between the components of the electric field parallel and perpendicular to its optic axis and rotates them at fixed positions separated by 22\degr.5. The HWP is followed by a WP which splits the light into two orthogonally polarised (extraordinary and ordinary) beams, separated by 22\arcsec\ (a Wollaston Mask is introduced before the HWP to avoid the overlapping of the two beams; Fig. 2.7 in FORS2 User manual). The last optical element before the detector is a grism (that may be preceded by an order-separating filter). The instrument design allows the user to implement a beam-swapping technique, in which two beams carrying light polarised in opposite directions are swapped during an observing series in which the HW is oriented at 0\degr, 22\fdg5, 45\degr, and 67\fdg5, \ldots\ with respect to the principal plane of the WP.  This allows one to largely suppress the effects of instrumental polarisation (e.g. \citealt{Bagnulo2009}). For our observations we used grism 300V with no order-separating filter, covering the spectral range 3,800\,\AA to 9,300\,\AA. We used a 1\farcs75 slit, for a spectral resolution of $\sim$250. The same instrument setup was adopted for all three observing epochs, with exposure times of 280\,s for the January and 300\,s for the February and March observations per single exposure, achieving the total exposure time of 2,240\,s and 2,400\,s respectively. Median seeing was 0\farcs66 for the January, 0\farcs58 for the February, and 0\farcs99 for the March observations.

\subsection{Data analysis  \label{sec:data}}

In all epochs, we obtained two sets of exposures ($N$ = 2 in Table \ref{t1}), that is, a total of eight fits files of sixteen light elements at the eight different HWP position angles from 0\degree\ to 157\fdg5 in 22\fdg5 intervals. A beam-swapping technique implemented took the first set at the HWP angle of 0\degree, 22\fdg5, 45\degree, and 67\fdg5, and the second set, which was 90\degree\ rotated from the first one, at the HWP angle of 90\degree, 112\fdg5, 135\degree, and 157\fdg5 in a row. Cosmic rays were removed using the L.A. Cosmic tool \citep{Dokkum2001}. Lamp calibration spectra (He+1, HgCd+2, Ar+1, and Ar+2) for wavelength calibration were taken under the identical observational configurations (Grism\_300V) after the target observations in 2 $\times$ 2 binning mode. We did not conduct flux calibration because we took ratios of the light elements to derive the Stokes parameters. Wavelength-calibrated spectra of each light component were then extracted using APALL in IRAF.  In APALL, the background of each spectrum was fitted by a linear function and then the spectrum was  traced by the third-order Chebyshev function. The resultant one-dimensional spectrum showed the root-mean-square (RMS) of $\sim$0.07--0.35 (pixel)$^{\rm -1}$. We repeated these extraction steps using six different aperture sizes (corresponding to 1,000 km--7,500 km in cometocentric distance).

The small RMS in the pre-processing stage confirmed that the subtraction of the slowly-varying background feature works well, while the small FoV of the slit (22\arcsec\ in width) limits the elimination of bright telluric lines, particularly $\sim$5,500, 6,300, 6,870, and 7,600 \AA\ for O$_{\rm 2}$, from the observed data. The remnants of the subtraction are discernible in the final spectrum (Figs. \ref{Fig02}--\ref{Fig04}), showing slight deviations with larger error bars than the ambient points.  We discarded the edge of the wavelength regions ($\lambda$ $<$ 4,000 \AA\ and $\lambda$ $>$ 9,000 \AA) that showed large fluctuations and binned the extracted data points to have a wavelength interval of $\sim$52 \AA. We found radial variations neither in count nor shape of the spectra obtained with different aperture sizes. (For instance, Figs. \ref{Fig12}--\ref{Fig14} in Appendix \ref{sec:app} show negligible differences between the spectra with the aperture sizes of 1,000 km and 7,500 km. Spectra with other aperture sizes also showed no difference within the uncertainties.) This is probably because the diffuse background coma structures visible in the sky images of Fig. \ref{Fig01} were too feeble to be observed with the exposure time applied in the PMOS mode, allowing us mainly to sample the central part of the coma of 67P. Since the larger the aperture, the more significant error becomes, we decided to analyse the spectrum extracted with the aperture size corresponding to 1,000 km in cometocentric distance.

Stokes parameters of the total intensity $I$ and the linear polarisation degree $Q$ and $U$ were derived using the difference method \citep{Bagnulo2009}. The reduced Stokes parameters $q$ and $u$ were calculated as 
\begin{equation}
q = \frac{Q}{I} = \frac{1}{2N} \sum_{i=1}^{N} \bigg[\bigg(\frac{f_{\rm ||} - f_{\rm \perp}}{f_{\rm ||} + f_{\rm \perp}} \bigg)_{X} - \bigg(\frac{f_{\rm ||} - f_{\rm \perp}}{f_{\rm ||} + f_{\rm \perp}} \bigg)_{X + 45\degree} \bigg]~,
\label{eq:eq1}
\end{equation}
and
\begin{equation}
u = \frac{U}{I} = \frac{1}{2N} \sum_{i=1}^{N} \bigg[\bigg(\frac{f_{\rm ||} - f_{\rm \perp}}{f_{\rm ||} + f_{\rm \perp}} \bigg)_{Y} - \bigg(\frac{f_{\rm ||} - f_{\rm \perp}}{f_{\rm ||} + f_{\rm \perp}} \bigg)_{Y + 45\degree} \bigg]~,
\label{eq:eq2}
\end{equation}
\noindent where $N$ is the number of pair of exposures ($N$ = 2 in this study); $f_{\rm ||}$ and $f_{\rm \perp}$ denote the photon counts of ordinary and extraordinary rays at each HWP position angle $\beta$; $\beta$ = $X$ $\in$ 0\degree\ and 90\degree\ for $q$ estimation, while $\beta$ = $Y$ $\in$ 22\fdg5 and 112\fdg5 for $u$ estimation; and Stokes $I$, the total intensity, can be expressed as
\begin{equation}
I = \frac{1}{4} \sum_{j=1}^{8} (f_{\rm ||} + f_{\rm \perp})_{j}~,
\label{eq:eq3}
\end{equation}
\noindent where $j$ is the eight HWP position angles of 0\degree, 22\fdg5, ..., 157\fdg5 in $\pi$/8 intervals.  Since we did not make absolute flux calibration in this study, the multiplicative factor (1/4) in Eq. \ref{eq:eq3} is of little importance. The usage of ratios between extraordinary and ordinary ray fluxes (split by the WP) in Eq. \ref{eq:eq1} allows us to remove instrumental effects such as flat fielding and the change in sky transparency between exposures \citep{Bagnulo2009}.

The derived reduced Stokes parameters were then corrected for the wavelength-dependent deviation of the HWP position angle from the nominal angle (i.e. chromatism; \citealt{Bagnulo2009}) using the equations of 
\begin{equation}
\begin{aligned}
\noindent q' &= q \cos 2\epsilon + u \sin 2\epsilon \\
u' &= -q \sin 2\epsilon + u \cos 2\epsilon~,
\end{aligned}
\label{eq:eq4}
\end{equation}
\noindent where $\epsilon$ is the deviation angle tabulated in the FORS2 user manual and the webpage\footnote{https://www.eso.org/sci/facilities/paranal/instruments/fors/inst/pola.html}. Next, we corrected for the wavelength-dependent instrumental polarisations $q_{\rm inst}$ and $u_{\rm inst}$ using Eq. 7 in \citet{Cikota2017} and converted the Stokes parameters in the instrument reference system to the celestial reference system as follows:
\begin{equation}
\begin{aligned}
q'' &= (q' - q_{\rm inst}) \cos 2\chi + (u' - u_{\rm inst}) \sin 2\chi \\
u'' &= -(q' - q_{\rm inst}) \sin 2\chi + (u' - u_{\rm inst}) \cos 2\chi~,
\label{eq:eq5}
\end{aligned}
\end{equation}
\noindent where $\chi$ denotes the WP position angle, counted counterclockwise from North to eastward, which corresponds to the header keyword of ADA.POSANG. The fraction of linear polarisation (i.e. the polarisation) and its position angle are then expressed as
\begin{equation}
P = \frac{1}{p_{\rm eff}} \sqrt{q''^{\rm 2} + u''^{\rm 2}}~,
\label{eq:eq6}
\end{equation}
and
\begin{equation}
\theta_{\rm P} = \frac{1}{2} \arctan \bigg(\frac{u''}{q''}\bigg)~,
\label{eq:eq7}
\end{equation}
\noindent where $p_{\rm eff}$ is the polarisation efficiency at each wavelength. Since the $p_{\rm eff}$ was measured at discrete wavelengths of 4,150--10,575 \AA, we interpolated the trend line of the $p_{\rm eff}$ to obtain the value at our wavelength positions.

In solar system science, the polarisation is usually expressed in a reference system whose reference direction is perpendicular to the scattering plane (a plane containing the Sun--comet--Earth). We thus transformed $P$ and $\theta_{\rm P}$ (measured with respect to the equatorial system) into the reference system using
\begin{equation}
P_{\rm r} = P \cos (2\theta_{\rm r})~,
\label{eq:eq8}
\end{equation}
and
\begin{equation}
\theta_{\rm r} = \theta_{\rm P} - \bigg(\phi \pm \frac{\pi}{2} \bigg)~,
\label{eq:eq9}
\end{equation}
\noindent where $\phi$ represents the position angle of the scattering plane, the sign of which ($\pm$ in Eq. \ref{eq:eq9}) was chosen to satisfy 0 $\leq$ ($\phi$ $\pm$ $\pi$/2) $\leq$ $\pi$ \citep{Chernova1993}. $\theta_{\rm r}$ can be used to check the reliability of the data as follows. In a random distribution of the coma dust, $\theta_{\rm r}$ $\sim$ 90\degree\ ($\sim$0\degree) is expected when the polarisation vector aligns perpendicular (parallel) to the scattering plane, and the resultant $P_{\rm r}$ becomes positive (negative). 
Uncertainties, including photon-noise on $P$ (Eq. 3 in \citealt{Bagnulo2017}) and the conversion to $P_{\rm r}$, were estimated by following standard error propagation methods. In our datasets, points free from the remnant of major telluric lines ($\sim$5,500, 6,300, 6,870, and 7,600 \AA\ for O$_{\rm 2}$) enabled us to retrieve reliable $P_{\rm r}$ values, showing the appropriate $\theta_{\rm r}$ alignments as expected.
\\

\section{Results \label{sec:res}}

\begin{figure}[!t]
\centering
\includegraphics[width=9cm]{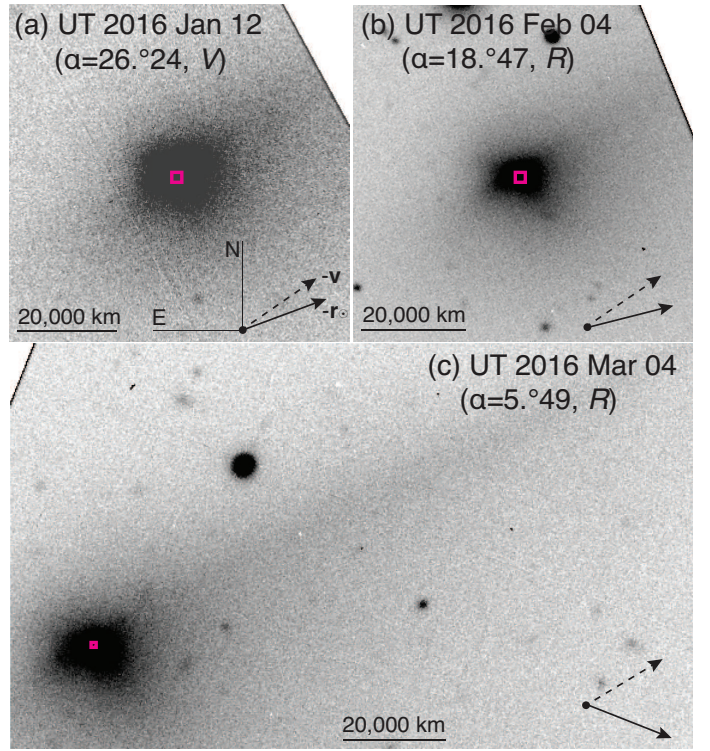}
\caption{Sky images of 67P obtained from a single exposure. The phase angle and equivalent filter range are provided on top of each figure. Panel (a) was obtained with an 120-sec exposure, while panels (b) and (c) were obtained with a 60-sec exposure. Panels (a) and (b) cover the FoV of 1\farcm0 $\times$ 1\farcm0, while panel (c) spans 2\farcm0 $\times$ 1\farcm2 to encompass the neck-line structure. North is up and east to the left. The level of brightness was inverted and arbitrarily adjusted to enhance weak coma structures. The 1,000-km-sized aperture applied in this study is overlaid as a magenta box, covering the inner coma part. The solid and dashed arrows denote the Sun--comet radius vectors and the comet's negative heliocentric velocity vectors, respectively. Background structures are either stars or cosmic rays.
} 
\label{Fig01}
\end{figure}

Fig. \ref{Fig01} shows sky images of 67P at three epochs obtained from a single exposure. 
The coma structure is highly asymmetric, showing southeastern and sunward jets with feeble dust tails both in the anti-solar ($-$\vec{r_{\rm \odot}}) and negative velocity ($-$\vec{v}) directions. The neck-line structure was visible at all epochs, but most clearly in March. It stretches up to $\sim$2.9 $\times$ 10$^{\rm 5}$ km in the sky plane beyond the figure, as confirmed by previous ground-based observations \citep{Snodgrass2017,Rosenbush2017,Knight2017}. In this section, we present the dependences of 67P's $P_{\rm r}$ (Eq. \ref{eq:eq8}) on wavelength (Sect. \ref{sec:res1}) and phase angle (Sect. \ref{sec:res2}), and the dependence of the intensity (Eq. \ref{eq:eq3}) on wavelength (i.e. colour; Sect. \ref{sec:res3}).

\subsection{Dependence of the polarisation degree on wavelength \label{sec:res1}}

\begin{figure}[!tb]
\centering
\includegraphics[width=9cm]{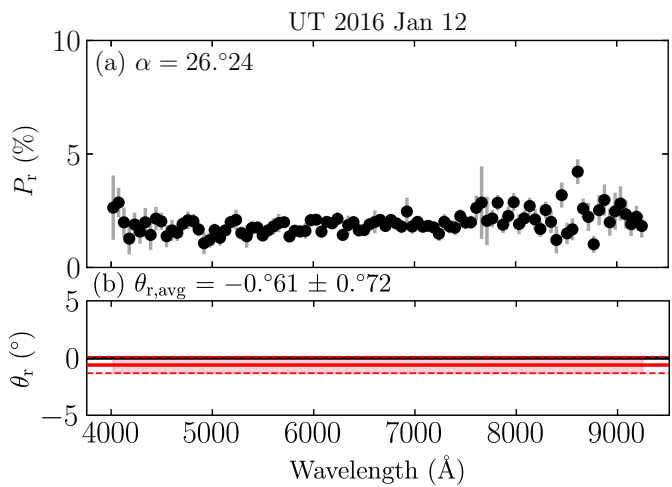}
\caption{$P_{\rm r}$ of 67P dust in \% (a) and the mean and 1-$\sigma$ error of the position angle of $P_{\rm r}$ with regard to the scattering plane ($\theta_{\rm r}$) as a function of wavelength (b) observed at $\alpha$ = 26\fdg4 on UT 2016 Jan 12. The black points in panel (a) are binned to have wavelength intervals of $\sim$52 \AA. The large error bars result from the remnants of the skylines that were not perfectly subtracted due to the limited slit width. The red solid and upper and lower dashed lines in panel (b) indicate the weighted mean of $\theta_{\rm r}$ ($\theta_{\rm r, avg}$) and its error (the standard deviation divided by the square of the number of the data points), respectively. The black line marks $\theta_{r}$ = 0\degree, the expected value when the dust is distributed randomly in the coma. } 
\label{Fig02}
\end{figure}

\begin{figure}[!htb]
\centering
\includegraphics[width=9cm]{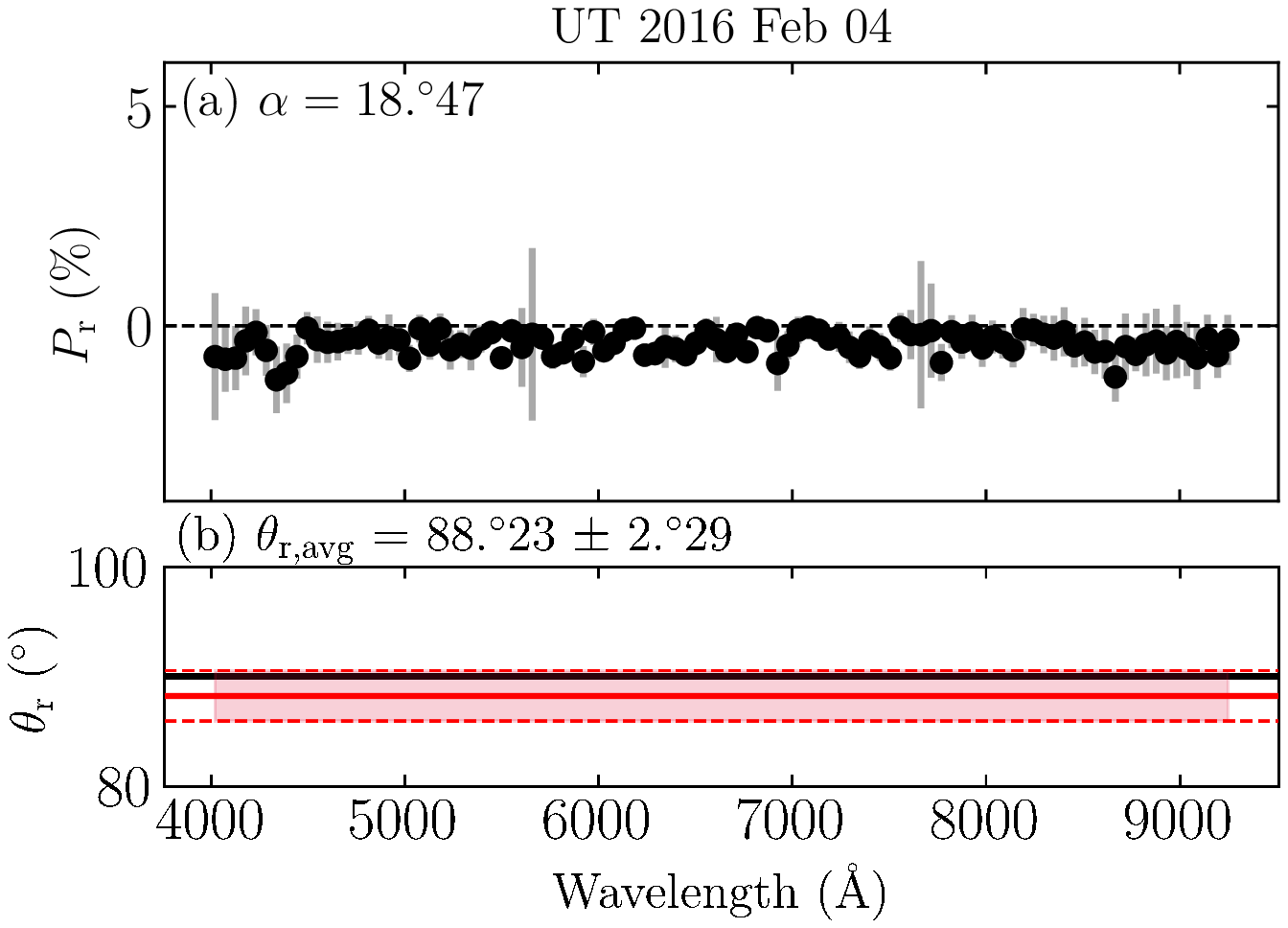}
\caption{Same as Figure \ref{Fig02}, but at $\alpha$ = 18\fdg47 on UT 2016 Feb 04. The black line indicates $\theta_{r}$ = 90\degree, the expected value when the dust is distributed randomly in the coma. } 
\label{Fig03}
\end{figure}

\begin{figure}[!htb]
\centering
\includegraphics[width=9cm]{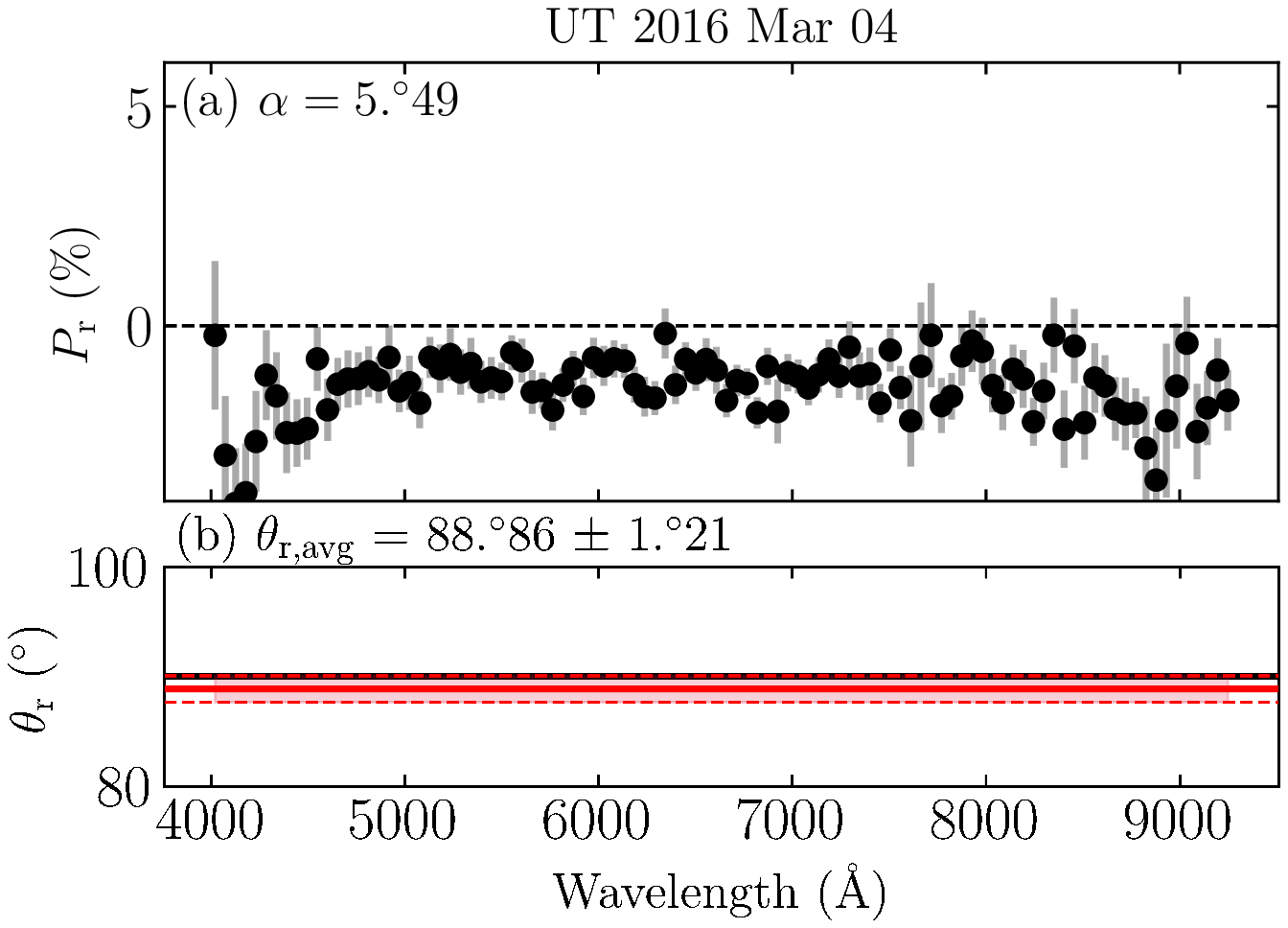}
\caption{Same as Figure \ref{Fig02}, but at $\alpha$ = 5\fdg49 on UT 2016 Mar 04. The black line indicates $\theta_{r}$ = 90\degree, the expected value when the dust is distributed randomly in the coma. }
\label{Fig04}
\end{figure}

Figs. \ref{Fig02}--\ref{Fig04} show the wavelength dependence of the polarisation degree, $P_{\rm r}(\lambda)$, of 67P dust on UT 2016 Jan 12, Feb 04, and Mar 04, respectively. Weighted mean and error of the $\theta_{\rm r}$ points over 4,000--9,000 \AA\ are also provided in panel (b) of each figure to check the reliability of data processing. In January (Fig. \ref{Fig02}), the dust's polarisation position angle is aligned to the perpendicular direction to the scattering plane ($\theta_{\rm r}$ $\sim$ 0\degree), such that the resultant $P_{\rm r}$ has a positive sign, while $P_{\rm r}$ of 67P dust in February and March (Figs. \ref{Fig03} and \ref{Fig04}) has a negative sign because of the polarisation vector along the scattering plane ($\theta_{\rm r}$ $\sim$ 90\degree). Except for several deviating points with large error bars due to the remnants of the sky emission lines (Sect. \ref{sec:data}), we do not find any gas emission line features at all epochs. This is consistent with the absence of a spherical gas coma structure in the sky images (Fig. \ref{Fig01}) and also with the results of the spectroscopic study using the VLT/FORS2 with the longer exposure time (300--600 sec) taken on UT 2016 Mar 04 \citep{Opitom2017}.  Although \citet{Opitom2020} later find the weak water-sublimation-driven [OI] emission at 6,300 \AA\ around the nucleus in 2016 March in VLT's integral field unit spectrograph (MUSE) data, this line is overlapped with the sky emission line, thus difficult for our data to discriminate the two lines.

In order to quantify the spectral dependence of $P_{\rm r}$, we make a linear least-square fitting. Since the comet's faintness makes the data points at both ends of the covered wavelength range fluctuating, we use only the intermediate points in the 4,550--7,560 \AA\ region for fitting and apply the solution to the whole wavelength range (4,000--9,000 \AA). The concept of `Polarimetric colour ($PC$)' is defined as
\begin{equation}
PC = \frac{\Delta P_{\rm r}}{\Delta \lambda} = \frac{P_{\rm r}(\lambda_{\rm 2}) - P_{\rm r}(\lambda_{\rm 1})}{\lambda_{\rm 2} - \lambda_{\rm 1}}~,
\label{eq:eq10}
\end{equation}
\noindent where $P_{\rm r}$($\lambda_{\rm 1}$) and $P_{\rm r}$($\lambda_{\rm 2}$) are the $P_{\rm r}$ values in percent measured at the wavelengths of $\lambda_{\rm 1}$ = 4,550 \AA\ and $\lambda_{\rm 2}$ = 7,560 \AA, respectively. Conventionally, a positive $PC$ is labeled as red, and a negative $PC$ is labeled as blue. With the linear equations of $a\cdot\lambda+b$, where $a$ and $b$ are the slope and y-intercept of the fitting function, we obtain the $PC$: (0.14 $\pm$ 0.08) \% (1,000 \AA)$^{\rm -1}$  for Fig. \ref{Fig02}; ($-$0.03 $\pm$ 0.07) \% (1,000 \AA)$^{\rm -1}$ for Fig. \ref{Fig03}; and (0.03 $\pm$ 0.12) \% (1,000 \AA)$^{\rm -1}$ for Fig. \ref{Fig04}.
Provided that we exclude the deviating data points on the short-end wavelength from consideration (particularly severe in March at $\lambda$ $<$ 5,000 \AA\ due to the low brightness of the comet), the {\it PC} values on either side of 5,500 \AA\ (the wavelength around which the dust colour changes its slope; Sect. \ref{sec:res3}) are consistent within the error ranges with those obtained from a total wavelength coverage.

\begin{figure}[!htb]
\centering
\includegraphics[width=9cm]{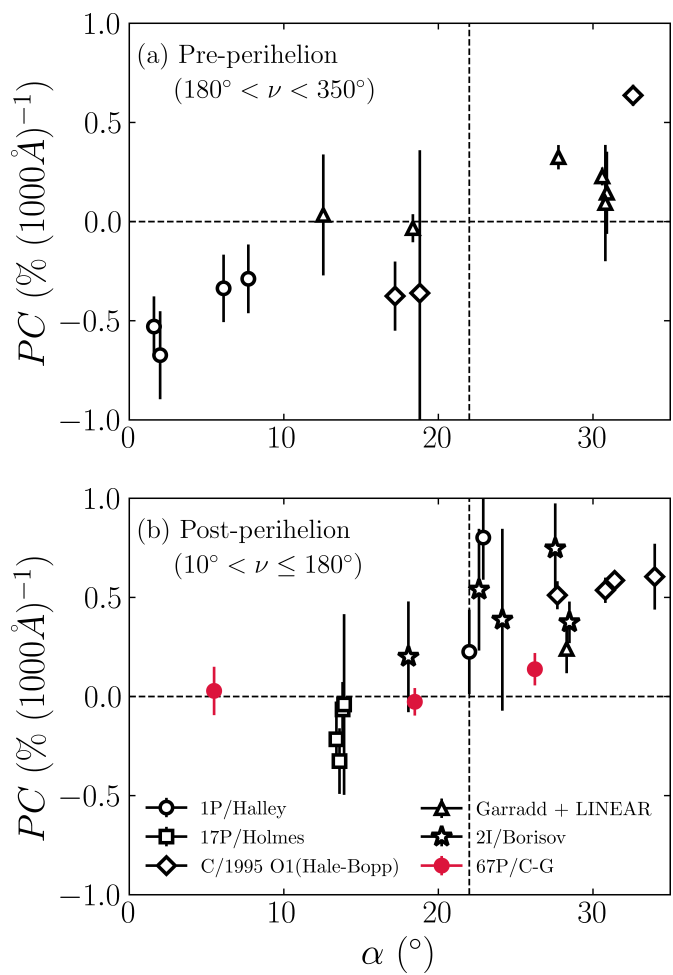}
\caption{The polarimetric colour ($PC$) of comets over the $\alpha$ range of 0\degree--35\degree. All data points except 67P in this study are quoted from the NASA/PDS comet polarimetric archive \citep{Kiselev2017} and \citet{Bagnulo2021}. Panels (a) and (b) present the data in pre- (180\degree\ $<$ $\nu$ $<$ 350\degree, where $\nu$ means the true anomaly) and post-perihelion (10\degree\ $<$ $\nu$ $\leq$ 180\degree), respectively. There are no available points around the perihelion (350\degree\ $\leq$ $\nu$ $\leq$ 10\degree). Red circles of 67P are obtained from the least-square linear fit of 67P dust (Figs. \ref{Fig02}--\ref{Fig04}), and their error bars indicate 1-$\sigma$ uncertainties of the fitting. Black open symbols are derived from the comet data observed by aperture polarimetric observations. Thus, the error bars of their $PC$s are derived from the error propagation of $P_{\rm r}$ values measured at discrete wavelength points. `Garradd` and `LINEAR` denote comets C/2009 P1 (Garradd) and C/2000 WM1 (LINEAR), respectively.}
\label{Fig05}
\end{figure}

Next, we compare 67P's spectral dependence of $P_{\rm r}$ with those of other comets taken at a similar range of $\alpha$. $PC$ of the cometary dust depends on $\alpha$. In general, for most comets, the $PC$ is red over the wavelength of interest at $\alpha$ > 25\degree, around which the $PC$ is relatively neutral but quickly increases the slope as $\alpha$ increases. The $PC$ at $\alpha$ < 20\degree\ tends to be slightly blue \citep{Kolokolova2004,Kiselev2015}. Fig. \ref{Fig05} shows the $P_{\rm r}$ distribution of cometary dust on the wavelength at the $\alpha$ range of 0\degree--35\degree\ adopted from the NASA/PDS comet polarimetric archive \citep{Kiselev2017} and \citet{Bagnulo2021}. We select data only taken with the narrow-band continuum filters and at multi-wavelength (quasi-)simultaneously (i.e. at least on the same day) to minimise the possible gas contamination and systematic difference induced by the nucleus rotation. Since the UC narrow-band filter (3,656/84 \AA) somewhat suffers from C$_{\rm 3}$ emission \citep{Kiselev1996}, we focus on the wavelength dependence in the range of $>$4,000 \AA.

In Fig. \ref{Fig05}, we can see a change of sign of $PC$ around the vertical dashed line from negative at $\alpha$ $\lesssim$ 22\degree\ (the negative polarisation branch; NPB) to positive at larger $\alpha$ (the positive polarisation branch; PPB), as expected \citep{Kiselev2015}. Both in pre- and post-perihelion phases, the comets, regardless of their dynamical groups, tend to follow this general trend. On the other hand, 67P dust shows a much flatter $PC$ both in the NPB and PPB than other comets, particularly compared to non- and long-period comets. 

\subsection{Dependence of the polarisation degree on phase angle \label{sec:res2}}

Low albedos and aggregate structures of cometary dust lead to a general $\alpha$ dependence of $P_{\rm r}$, parameterised by a shallow NPB with an average minimum polarisation $P_{\rm min}$ $\approx$ $-$1.5 \% at $\alpha_{\rm min}$ $\approx$ 10\degree, and an inversion angle where $P_{\rm r}$ changes its sign from negative to positive at $\alpha_{\rm 0}$ $\approx$ 22\degree, and a bell-shaped PPB with a maximum polarisation $P_{\rm max}$ $\approx$ 25--30 \% at $\alpha_{\rm max}$ $\sim$ 95\degree\ in the wavelength region we are interested in \citep{Kiselev2015}. The most probable explanation for the change of $P_{\rm r}$'s sign is that the interference effects between the multiple-scattered light around the back-scattering region yields the change of the alignment of the polarisation vector parallel to the scattering plane \citep{Muinonen2015}.

\begin{figure}[!htb]
\centering
\includegraphics[width=9cm]{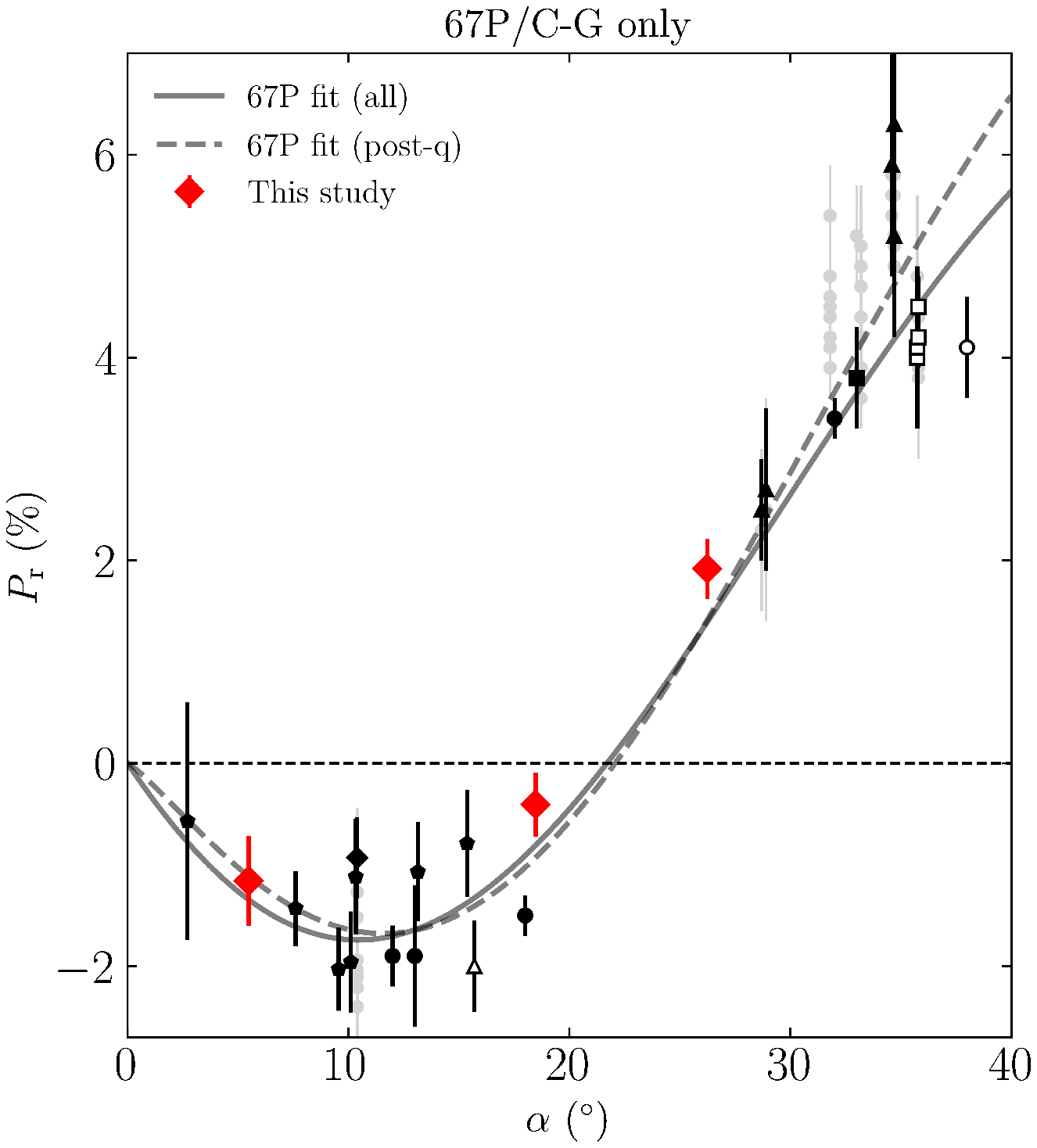}
\caption{The polarisation of 67P dust versus phase angle. Open and filled symbols denote the pre- and post-perihelion data, respectively. Black symbols are the archival data \citep{Kiselev2017} where different shapes indicate the datasets obtained by different works: circle \citep{Myers1984}, square \citep{Hadamcik2016}, triangle \citep{Hadamcik2010}, polygon \citep{Stinson2016}, and diamond \citep{Rosenbush2017}. Red diamonds denote our spectropolarimetric data. Background grey symbols \citep{Kiselev2017} are excluded from the analysis due to the significant difference in the aperture sizes from ours. The solid and dashed curves indicate the average dependence derived from Eq. \ref{eq:eq11} for the data as a whole and in post-perihelion, respectively. The best-fit parameters of the curves are provided in the text.}
\label{Fig06}
\end{figure}

We first compile all $P_{\rm r}$ of 67P dust from the NASA/PDS comet polarimetric archive \citep{Kiselev2017} to compare the phase angle dependence with our data. Given that the majority of the archival data were taken in the Red domain (central wavelength of 6,200--7,300 \AA) and that the highest S/N in our data was achieved in the 5,000--7,000 \AA\ range with flat wavelength dependence (Figs. \ref{Fig02}--\ref{Fig04}), we decide to utilise the archival data taken in the Red domain. We thus take the weighted mean of our data points over 5,000--7,000 \AA\ at each epoch and use the result as a nominal value. Only narrow-band $P_{\rm r}$ and the data proven to be free from gas contamination in the Red domain are selected from the archive for the analysis. The archival data having either very small or large apertures are excluded. To minimise possible dependence of $P_{\rm r}$ on the aperture size, we adopt the data points having the aperture size in the same order of magnitude as ours (1,000 km in radius), resulting in the range of 1,000--10,000 km cometocentric distance (on average of $\sim$2,300 km). The dataset with the 10,000-km-sized aperture comes from \citet{Stinson2016}, where the comet was in its outbound orbit, showing weak but stable coma activity. The large scatter of their data across the fitted line seems to introduce no bias (black circles in Fig. \ref{Fig06}), such that we decide to use them. The average $P_{\rm r}$ dependence on the phase angle, $P_{\rm r}(\alpha)$, of the selected data is then obtained using the empirical trigonometric function of \citet{Lumme1993}: 
\begin{equation}
 P(\alpha) = b~\sin^{c_{\rm 1}}(\alpha) \cos^{c_{\rm 2}} \bigg(\frac{\alpha}{2}\bigg)  \sin(\alpha - \alpha_{\rm 0})~,
\label{eq:eq11}
\end{equation}
\noindent where $b$, $c_{\rm 1}$, $c_{\rm 2}$, and $\alpha_{\rm 0}$ are the wavelength-dependent free parameters shaping the curve.

Fig. \ref{Fig06} shows the resultant $P_{\rm r}(\alpha)$ of 67P dust. The number of pre-perihelion data (open symbols) of 67P is insufficient to fit the trend, and thus the trends for the data as a whole (solid curve) and those in post-perihelion data (dashed curve) are examined. The best-fit parameters of the curve for all 67P data are $b$ = 52.32 $\pm$ 20.87 \%, $c_{\rm 1}$ = 1.01 $\pm$ 0.24, $c_{\rm 2}$ = 9.99 $\pm$ 5.82, and $\alpha_{\rm 0}$ = 21\fdg71 $\pm$ 0\fdg51. Those for 67P post-perihelion data are $b$ = 68.10 $\pm$ 40.10 \%, $c_{\rm 1}$ = 1.21 $\pm$ 0.32, $c_{\rm 2}$ = 9.99 $\pm$ 10.32, and $\alpha_{\rm 0}$ = 22\fdg07 $\pm$ 0\fdg57. Since the number of 67P data is small and they cover $\alpha$ $<$ 40\degree, the uncertainty of the fitting parameter $b$, which is related to the amplitude of the curves (i.e. the maximum $P_{\rm r}$ at $\alpha$ $\sim$ 95\degree), becomes significant. The result shows that the pre-perihelion data points appear to be located in the lower side of the average trend both in the NPB and PPB. \citet{Hadamcik2016} attribute such a systematic $P_{\rm r}(\alpha)$ difference of 67P dust between the pre- and post-perihelion to the different characteristics of the dust. Although the small number of pre-perihelion data prevents us from making in-depth analysis, we confirm the consistently lower position of the pre-perihelion data with regard to the fitted curves, supporting the evolution of the dust properties throughout the perihelion passage. 

\begin{figure}[!htb]
\centering
\includegraphics[width=9cm]{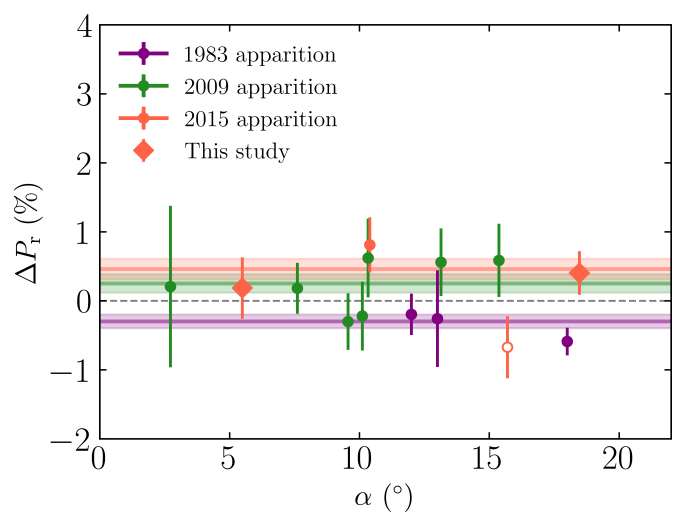}
\caption{The departure of the $P_{\rm r}$ of 67P dust in each apparition from to the average trend. The black fitting line in Fig. \ref{Fig06} now corresponds to the horizontal dashed lines at $\Delta P_{\rm r}$ = 0. The horizontal solid lines and shaded areas denote the weighted mean and error, in which the colour scheme follows the symbols in legend. Open and filled symbols denote the pre- and post-perihelion data of 67P dust. Only post-perihelion data were used to estimate $\Delta P_{\rm r}$. }
\label{Fig07}
\end{figure}

We also investigate how the $P_{\rm r}(\alpha)$ varies with time. We subtract the fitted $P_{\rm r}$ at the given $\alpha$ from the observed ($\Delta P_{\rm r}$ = $P_{\rm r, obs}$ $-$ $P_{\rm r, fitted}$) and take the weighted mean of the $\Delta P_{\rm r}$ in the same apparition. Since it is better to analyse the data in the NPB separately from those in the PPB \citep{Muinonen2015}, we delimit the $\alpha$ range of $<$22\degree\ to focus on the NPB. Fig. \ref{Fig07} shows the departure of the $P_{\rm r}$ of 67P dust ($\Delta P_{\rm r}$) in different apparitions from the general trend of the phase curve (i.e. the solid line in Fig. \ref{Fig06} now equals $\Delta P_{\rm r}$ = 0). The $\Delta P_{\rm r}$ and its error are ($-$0.30 $\pm$ 0.10) \%, (0.25 $\pm$ 0.13) \%, and (0.46 $\pm$ 0.15) \% for the 1983, 2009, and 2015 apparitions, respectively. We do not include the pre-perihelion point (open orange circle) for the retrieval. Although the $P_{\rm r}(\alpha)$ difference between the last two apparitions is statistically insignificant, we find an overall decrease in the absolute $P_{\rm r}$ values in the NPB (i.e. a decrease in the depth of the NPB) as the apparition proceeds. 

\begin{figure}[!htb]
\centering
\includegraphics[width=9cm]{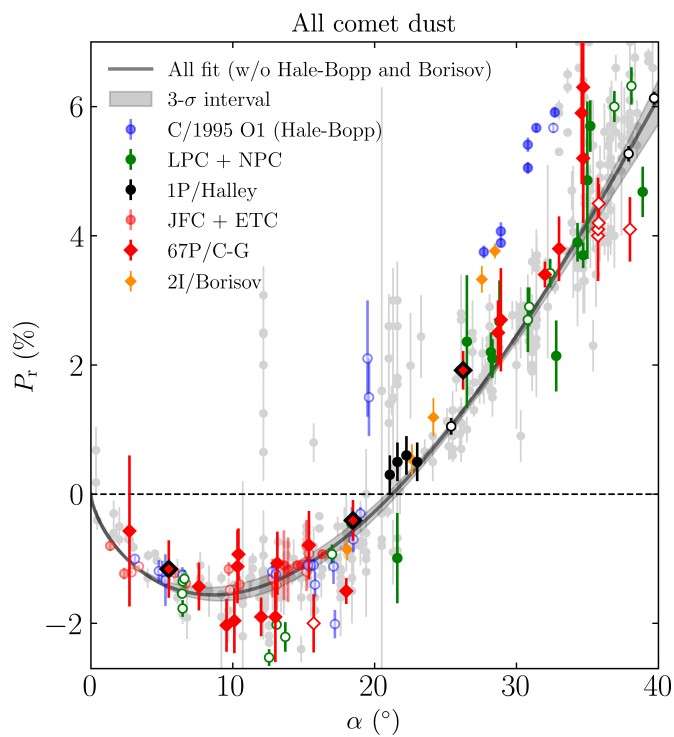}
\caption{The polarisation of all available comets' dust versus phase angle. The black solid line and shaded areas denote the average trend of the dependence and its 3-$\sigma$ error, respectively. As in Fig. \ref{Fig06}, open and filled symbols denote the pre- and post-perihelion data, while background grey circles represent data points discarded from the fitting because of either their large aperture size ($\gg$10,000 km in cometocentric distance) and/or the possible gas contamination. All symbols are the archival data of \citet{Kiselev2017}, except for the four points of 2I/Borisov \citep{Bagnulo2021} and three points of 67P (this study). We exclude unique comets C/1995 O1 (Hale-Bopp) and 2I/Borisov from the fitting, since their $P_{\rm r}$ has been found to be significantly higher. The red diamonds edged with black lines are our spectropolarimetric data. The best fit parameters of the fitted curve are $b$ = 25.69 $\pm$ 1.27 \%, $c_{\rm 1}$ = 0.67 $\pm$ 0.02, $c_{\rm 2}$ = (1.54 $\pm$ 0.85) $\times$ 10$^{\rm -13}$, and $\alpha_{\rm 0}$ = 21\fdg25 $\pm$ 0\fdg14. Detailed description of the legend is provided in the text.}
\label{Fig08}
\end{figure}

Finally, we compare the $P_{\rm r}(\alpha)$ of dust between 67P and other comets. The narrow-band data (or broad-band ones unambiguously proved to be free from gas contamination) taken in the Red domain from the NASA/PDS archive \citep{Kiselev2017} and the data of an extrasolar comet, 2I/Borisov, taken in $R_{\rm F}$ filter (6,550/1,650 \AA; \citealt{Bagnulo2021}) are used for this comparative study. Again, the average trend is estimated from Eq. \ref{eq:eq11} for the data whose aperture size ranges from $\sim$1,000--10,000 km (on average $\sim$4,000 km in cometocentric distance). We exclude unique comets C/1995 O1 (Hale-Bopp) and 2I/Borisov from the fitting due to their significant $P_{\rm r}$ excess compared to the majority of comets \citep{Hadamcik2003,Kikuchi2006,Bagnulo2021} and the large aperture size (for Hale-Bopp). Fig. \ref{Fig08} shows the results of the compilation. The LPC and NPC stand for long- and non-period comets, respectively, in which comets used are C/1990 K1 (Levy), C/1996 B2 (Hyakutake), C/2000 WM1 (LINEAR), C/2001 A2 (LINEAR), C/2009 P1 (Garradd), and C/2010 R1 (LINEAR). The JFC and ETC stand for Jupiter-family and Encke-type comets, respectively, and used comets are 17P/Holmes, 67P, 74P/Smirnova-Chernykh, 78P/Gehrels, 152P/Helin-Lawrence, and 290P/Jager. The points of 67P are used for fitting but shown separately (as red diamonds) for comparison with other comets. The comet shows a similar $P_{\rm r}$ distribution with other JFC and ETC within the measurement uncertainties.

Most comets show deviations from the average to both directions, being on the upper and lower parts of the fitted curve. Such dispersion is partially a result of an unstable state in the cometary activity but would also be the consequence of observations made by various teams in slightly different wavelength domains (e.g. \citealt{Lasue2009}).
However, C/1995 O1 (Hale-Bopp) around its perihelion (25\degree\ $<$ $\alpha$ $<$ 35\degree) always shows excessively high $P_{\rm r}$ values compared to the majority of comets. $P_{\rm r}$ of 2I/Borisov shows a typical value in the NPB but as high as those of Hale-Bopp in the PPB \citep{Bagnulo2021}. For other comets, we do not find such a significant difference. In the NPB, the short-period comets (i.e. JFC and ETC) tend to extend to the slightly smaller absolute $P_{\rm r}$ values than LPC and NPC. The Oort-cloud origin comets tend to spread toward more negative $P_{\rm r}$ values.

\subsection{Dust colour \label{sec:res3}}

In addition to the $PC$, we can also obtain the wavelength dependence of the intensity (i.e. dust colour) by dividing the Stokes $I$ of the dust (Eq. \ref{eq:eq3}) by that of the solar spectrum. To this end, we observed a solar analog HD 67010 \citep{Datson2015} under the identical optics to the target observations (PMOS mode), with the HWP angle of 0\degree\ on UT 2016 February 16. Extraction and binning of the stellar spectrum performed in the same way as described in Sect. \ref{sec:obsdata}, yielding the extraction RMS of $\sim$0.02--0.13 (pixels)$^{\rm -1}$.  Assuming that the star is unpolarised, we obtain its intensity by adding the counts of $f_{\rm ||}$ and $f_{\rm \perp}$ and multiplying the sum by 2. A difference in airmass between the solar analog (on average 2.59) and 67P observations (Table \ref{t1}) can be compensated using 
\begin{equation}
\begin{aligned}
\noindent -2.5 \log (F_{\rm c, 0}) &= -2.5 \log (F_{\rm c}) - kX_{\rm c} + Z_{\rm c} \\
-2.5 \log (F_{\rm *, 0}) &= -2.5 \log (F_{\rm *}) - kX_{\rm *} + Z_{\rm *}~,
\end{aligned}
\label{eq:eq12}
\end{equation}
\noindent where the subscripts $c$ and $*$ indicate a quantity for the comet and the star. $F_{\rm c, 0}$ and $F_{\rm *, 0}$ are the fluxes of 67P dust and the solar analog outside atmosphere, respectively, and $F_{\rm c}$ and $F_{\rm *}$ are the counterparts measured at the airmass $X_{\rm c}$ ($I$ in Eq. \ref{eq:eq3}) and $X_{\rm *}$ (2.59). $k$ denotes the wavelength-dependent extinction coefficient. Assuming that the zero points on the observing nights of 67P (Z$_{\rm c}$) and the star (Z$_{\rm *}$) are roughly the same, Eq. \ref{eq:eq12} reduces to 
\begin{equation}
\bigg(\frac{F_{\rm c,0}}{F_{\rm *,0}}\bigg) = \bigg(\frac{F_{\rm c}}{F_{\rm *}}\bigg) \times 10^{k \rm (X_{\rm c} - X_{\rm *})/2.5}~,
\label{eq:eq13}
\end{equation}
\noindent the left side of which is the dust colour. For the $k$ values, we refer to the Paranal coefficients published by \citet{Patat2011}. The zero points measured in the B, V, and R filters at the four observing epochs are in fact slightly different within $\sim$0.01 mag. This introduces an uncertainty of $\lesssim$3 \% in comparing the relative brightness measured on different nights.

\begin{figure}[!htb]
\centering
\includegraphics[width=9cm]{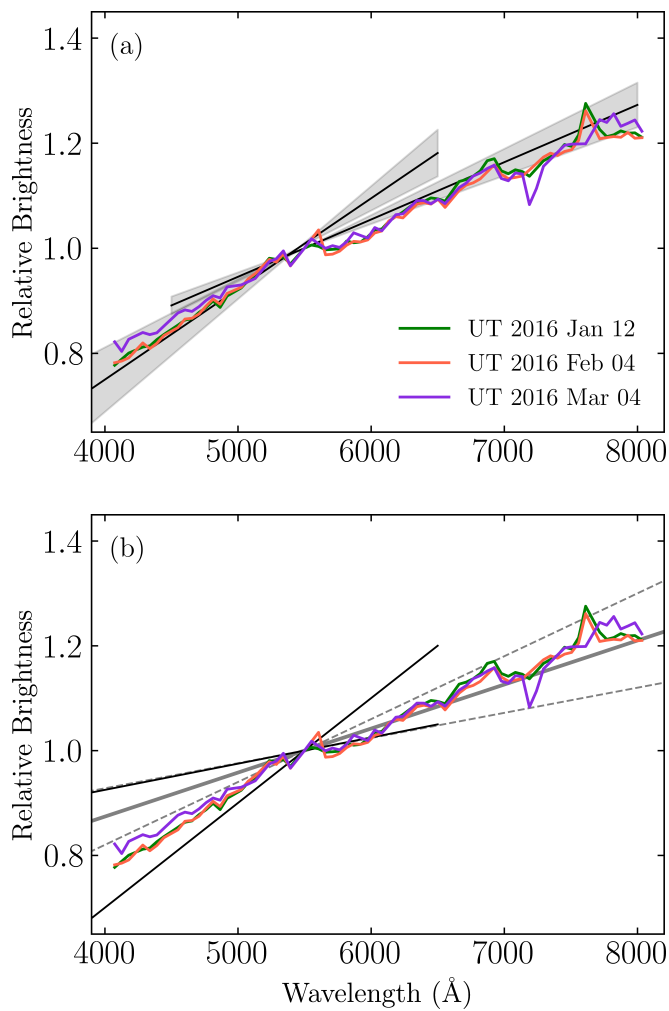}
\caption{The relative brightness of 67P dust normalised at 5,500 \AA. (a) The dust colour slopes (Eq. \ref{eq:eq14}) are shown as the solid lines with 3-$\sigma$ fitting errors in grey regions. (b) Two solid black lines indicate the lower (5 $\pm$ 2 \% (1,000 \AA)$^{\rm -1}$) and upper (18 $\pm$ 2 \% (1,000 \AA)$^{\rm -1}$) reflectivity limits of the dust of nine comets over 3,500--6,500 \AA, respectively \citep{Jewitt1986}. The thick grey solid and dashed lines indicate the average reflectivity trend and standard deviation of 31 active comets (8.4 $\pm$ 3.6 \% (1,000 \AA)$^{\rm -1}$) over 3,540--9,050 \AA\ used in \citet{Solontoi2012}. } 
\label{Fig09}
\end{figure}

Fig. \ref{Fig09}a shows the resultant relative brightness distribution of 67P dust over 4,000--8,000 \AA. Although incomplete correction of telluric absorption lines, particularly O$_{\rm 2}$ and H$_{\rm 2}$O lines, leaves bumpy structures at the longward of $\sim$6,500 \AA\ at all epochs, the dust continuum colour compensated for the atmospheric effects remains nearly the same over the two months. A change in the slope of the colour (i.e. the inflection point) appears around 5,500 \AA, consistent with the previous studies \citep{Frattin2017,Filacchione2020,Opitom2020}. To compare the colour presented here with other comets, we derive the normalised reflectivity $S'$ in the unit of \% (1,000 \AA)$^{\rm -1}$ using 
\begin{equation}
S' (\lambda_{\rm 1}, \lambda_{\rm 2}) = \frac{1}{S_{\rm mean}} \Bigg(\frac{S_{\lambda_{\rm 2}} - S_{\lambda_{\rm 1}}}{\lambda_{\rm 2} - \lambda_{\rm 1}}\Bigg)~
\label{eq:eq14}
\end{equation}
\noindent \citep{Jewitt1986}, where $\lambda_{\rm 1}$ and $\lambda_{\rm 2}$ are the short-end and long-end wavelengths considered. We set $\lambda_{\rm 1}$ = 4,000 \AA\ and $\lambda_{\rm 2}$ = 5,500 \AA\ for the shortward of the inflection point (``blue'' side), and $\lambda_{\rm 1}$ = 5,600 \AA\ and $\lambda_{\rm 2}$ = 7,900 \AA\ for the longward of the inflection point (``red'' side). $S_{\lambda_{\rm 1}}$, $S_{\lambda_{\rm 2}}$, and $S_{\rm mean}$ denote the relative brightness at $\lambda_{\rm 1}$ and $\lambda_{\rm 2}$ and the mean relative brightness in the wavelength range considered, respectively. Three colours in Fig. \ref{Fig09}a are combined into the average, yielding (17.3 $\pm$ 1.4) \% (1,000 \AA)$^{\rm -1}$ for the blue side and (10.9 $\pm$ 0.6) \% (1,000 \AA)$^{\rm -1}$ for the red side.

The $S'$ of 67P dust in Fig. \ref{Fig09}b is in line with the results of previous studies  \citep{Storrs1992,Schleicher2006,Hadamcik2009,Snodgrass2016,Rosenbush2017} that show slightly larger (i.e. redder colour) but still broadly consistent $S'$ of 67P dust with the typical colour range of short-period comets. 
\citet{Grun2016} report a three-hour-long outburst event accompanied by factors $>$100 increase of the near-nucleus coma brightness when Rosetta was at an altitude of 34 km on UT 2016 February 19. However, any significant aftermath to the event seems invisible in our colour data. 
Compared to variations of the brightness and the $Af\rho$ parameter (a proxy of the dust mass-loss rate of a comet; \citealt{A'Hearn1984}) from photometry showing a slight increase after the outburst on UT 2016 February 19 \citep{Boehnhardt2016,Grun2016}, the stable colour trend shown in our data would indicate that the dust properties over the two months remain more or less constant.
\\

\section{Discussion \label{sec:discuss}}

In this section, we describe the dust environment of 67P based on the results in Sect. \ref{sec:res}, in conjunction with the observations obtained from in situ and previous ground-based studies. We first constrain the size and structure of 67P dust during this period (Sect. \ref{sec:dis1}) and the possible amount of embedded water ice and the refractory-to-ice ratio of the dust particles (Sect. \ref{sec:dis2}).

\subsection{Constraints on the properties of 67P dust \label{sec:dis1}}

This study covers two months on the brink of the second equinox of 67P (at $r_{\rm H}$ = 2.63 au on UT 2016 March 21). The sub-solar latitude of the nucleus during the time of our observations varied from about $-$13\degree\ to $-$1\degree\ \citep{Preusker2017}, in which the comet's dust activity had already passed its peak and decreased smoothly \citep{Knight2017}. In the following, we suggest the most plausible properties of the dust attributable to our results in Sect. \ref{sec:res} by considering three main aspects --- the composition, size, and porosity of the dust.

\subsubsection{Conjecture based on the observational evidence in this study}

The primary constituents of cometary dust are silicates (mainly in the form of olivine and pyroxene), carbonaceous materials (amorphous carbon and organics), Fe-bearing sulfides, and ice \citep{Levasseur-Regourd2018}. Our colour data showing the broad consistency with active short-period comets (Fig. \ref{Fig09}b) would indicate not much difference of 67P dust from the typical dust in compositional terms of view. However, its slightly redder colour than the average would allow us to rule out the predominance of materials whose transparency decreases with wavelengths (e.g. water ice) in the dust particles \citep{Kolokolova1997b}. The polarimetric phase curve of 67P dust in the Red domain, which is consistent with the trend of most comets but shows a slightly shallower depth of the NPB than those of non- and long-period comets (Fig. \ref{Fig08}) and 67P in previous apparitions (Fig. \ref{Fig07}), support the ordinary composition of 67P dust dominated by absorbing materials \citep{Petrova2001,Hadamcik2009}. A large imaginary part of the complex refractive index (i.e. high absorptivity) can also explain the neutral $PC$ of 67P dust (Fig. \ref{Fig05}) that is flatter than those of other comets in similar observing geometry \citep{Kolokolova1997b}.

In terms of the dust's physical attributes, the dominance of large{\footnote{Dust with a size of $\gtrsim$tens of micrometres safely in the geometrical optics regime in the optical. This scattering regime is defined by $X$ $\gg$ 1, where $X$ := 2$\pi a$/$\lambda$ is the size parameter, $a$ is the radius of the dust, and $\lambda$ is the wavelength considered.}} over small dust{\footnote{Dust with a size of the order of $\sim$0.1--1 micrometres. This size range corresponds to the Rayleigh regime ($X$ $\ll$ 1) and Rayleigh-like small Mie regime ($X$ $\sim$ 1) in the optical.}}  can explain such red colour \citep{Kolokolova1997}, small $PC$ values \citep{Gustafson1999}, and shallower NPB depth \citep{Hadamcik2009} of 67P dust. All the results are also compatible with a low level of dust porosity. Laboratory experiments and computer modelling of cometary dust have shown that as the packing density of the cometary dust increases, the depth of the NPB becomes shallower \citep{Petrova2001,Hadamcik2009,Kolokolova2015}, the dust colour reddens \citep{Gustafson1999}, and the $PC$ becomes less red \citep{Gustafson1999}.

Taken as a whole, observations of 67P dust obtained with the 1,000 km-sized aperture in this study could be ascribed to the dominance of coarse ($\gtrsim$tens of micrometres in size), rather compact dust particles in the coma, whose light scattering characteristics are similar to those of the evolved near-surface dust layer likely common for short-period comets (e.g. \citealt{Hadamcik2009,Biele2015,Spohn2015,Kwon2018}). The Comet Nucleus Sounding Experiment by Radiowave Transmission (CONSERT), onboard Rosetta and its Philae lander,  indeed provides the evidence of a shallow subsurface of about 20 m depth, less porous and thus denser than the interior of the nucleus \citep{Kofman2020}.

\subsubsection{Comparison with the Rosetta observations and estimation of the plausible dust porosity}

In early 2016, mainly two different types of dust particles would coexist in the sampled 1,000-km-sized coma: (i) particles directly ejected from the nucleus; and (ii) particles which were ejected and had either fallen back to the nucleus surface, followed by re-ejection around perihelion, or remained in bound orbits without fallback \citep{Agarwal2016,Longobardo2020}. Type (ii) dust particles are primarily decimetre-sized aggregates \citep{Mottola2015,Agarwal2016,Pajola2017}, and their large size and compactness make them less sensitive to the gas drag and solar radiation pressure than fluffier dust counterparts \citep{Gundlach2020,Fulle2020}. They tend to linger in the circumnuclear coma of 67P ($<$100 km; e.g. \citealt{Bertini2019}), and we expect that their contribution to the scattering cross-section in the 1,000 km aperture would be small. For Type (i) dust particles, Rosetta observations classify them into two families --- compact and fluffy dust aggregates. Although the definition of the classification is slightly different between the Grain Impact Analyser and Dust Accumulator (GIADA) \citep{DellaCorte2015} and the Cometary Secondary Ion Mass Analyzer (COSIMA) \citep{Hornung2016,Langevin2016}, they show in common that fluffy dust particles have a fractal dimension of $<$2 and are prone to be fragmented, while compact counterparts have a porosity below 90 \% with higher tensile strengths and density \citep{Levasseur-Regourd2018}. The former prevails in the dust coma around perihelion, but contributes less beyond the outbound distance of $\sim$2 au \citep{Longobardo2020}. Relatively compact dust aggregates would then become the dominant signal in the background coma at the epochs we covered, showing no significant fragmentation as they move out of the nucleus at least in the $<$500 km distance range (\citealt{Frattin2021} and references therein).

In a synthesis paper of 67P dust, \citet{Guttler2019} classify dust particles collected throughout the mission depending on the size, porosity, and strength, and confirm the dominance of the role of `porous dust agglomerates' in the observed light scattering phenomena. This type of dust has a typical size of $\sim$100 $\mu$m with the porosity ranging from 10 \% to 95 \% (\citealt{Guttler2019} and references therein).
A majority of compact dust aggregates, which are initially named by COSIMA and GIADA, belong to this dominant group based on their physical properties. Therefore, we use the terminology of `porous dust agglomerates' in the following instead of compact dust aggregates, to demonstrate the dust with a lower porosity than fluffy dust aggregates. The composition of 67P dust in such a size scale shows similarity to a large extent with other cometary dust probed by past space missions, in which Mg-rich silicates and carbonaceous materials account for the majority of refractory components \citep{Levasseur-Regourd2018}.

In order to understand how porosity influences the polarimetric behaviour of such dust particles, we numerically simulate the light scattering of the dust with a representative size of 100 $\mu$m in radius and  three different cases of the dust porosity: 60 \%, 80 \%, and 95 \%. The first porosity value considered represents the solid-like (non-fragmenting) dust \citep{Hornung2016}, the second one is for the dust with an intermediate porosity, and the last one is for the fluffiest particle in the `porous dust agglomerate' regime \citep{Guttler2019}. 
We used the radiative transfer with reciprocal transactions method (R2T2) to calculate the scattering properties. R2T2 is an extension of the radiative transfer solution to the dense discrete random medium, including the coherent backscattering mechanism \citep{Muinonen2018,Vaisanen2019}. 
For the sake of simplicity, we do not consider the hierarchical nature of dust agglomerates (i.e. the assemblage of multiple clusters of smaller aggregates) but instead a single spherical aggregate structure. 
The modelled dust consists of spherical monomers{\footnote{More realistic monomers in shape would be non-spherical. However, we found that the outcome of light scattering computation using the polydisperse distribution of irregular monomers seems largely similar to the one using spherical monomers.}} following a power-law size distribution with the differential index of $-$3. Their minimum and maximum sizes are 0.05 $\mu$m and 0.25 $\mu$m, respectively. The modelled dust is constructed with 75 vol.\% of amorphous carbon \citep{Jager1998} and 25 vol.\% of Mg-rich amorphous pyroxene (Mg$_{\rm 0.8}$Fe$_{\rm 0.2}$SiO$_{\rm 3}$; \citealt{Dorschner1995}) to make dust albedo consistent with typical Halley-type dust composition \citep{Mann2004,Kimura2006}. To see the influence of the composition, we also test a case with 65 vol.\% and 35 vol.\% of the former and latter ingredients, respectively, also in agreement with overall results from 67P local studies \citep{Levasseur-Regourd2018}.
Each monomer has a single composition (either amorphous carbon or silicate). Although \citet{Bardyn2017} report carbon-rich dust particles of 67P whose refractory components would be dominated by organic matter in mass, its exact stoichiometry (especially for aliphatic organics; \citealt{Raponi2020}), let alone imaginary part of the complex refractive index across the broad wavelength range covered by our observations, is not yet available. As such, we select amorphous carbon to represent the organics to match the albedo of the modelled dust to the dark level of the observed dust particles. More detailed modelling in the compositional and structural aspects of the dust will be introduced in our future work, so the modelling in this paper aims only to illustrate the influence of the porosity on polarimetric behaviours of the dust particles and narrow down the most likely porosity values able to explain our results.

\begin{figure}[!htb]
\centering
\includegraphics[width=9cm]{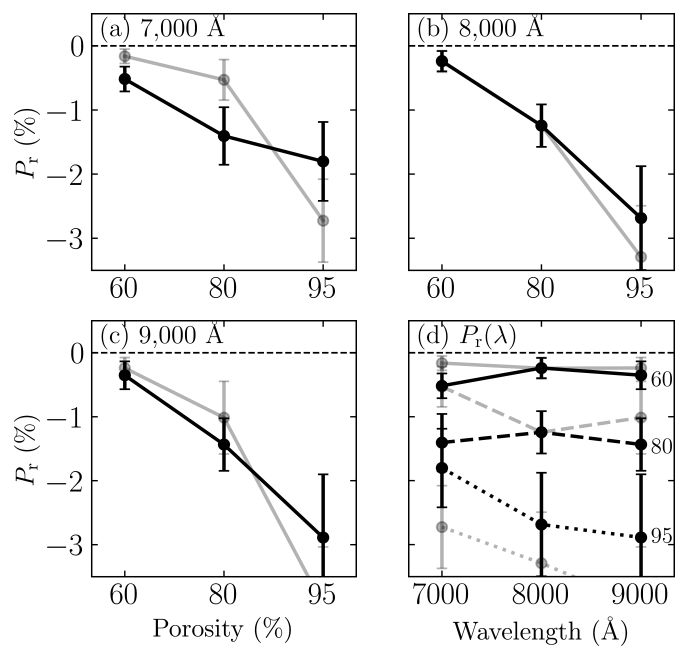}
\caption{$P_{\rm r}$ in the negative polarisation branch of the modelled dust aggregate having porosities of 60 \%, 80 \%, and 95 \% at the wavelengths of (a) 7,000 \AA, (b) 8,000 \AA, and (c) 9,000 \AA, and (d) the spectral dependence of $P_{\rm r}$ over the three wavelength domains. The results represent the mean $P_{\rm r}$ values and their standard deviations over the 1\degree--10\degree\ phase angle interval. Black (grey) symbols indicate the modelled results of the dust in the composition of 75 (65) vol.\% amorphous carbon + 25 (35) vol.\% silicate. A more detailed description of the modelling is provided in the text. The $P_{\rm r}$ of the transparent dust at 9,000 \AA\ is $-$3.94 \% and is not shown in this figure.} 
\label{Fig10}
\end{figure}

Figure \ref{Fig10} summarises the modelling results. Changes in $P_{\rm r}$ of the dust particles with regard to their porosity and wavelengths are shown. Black (grey) symbols indicate the modelled results with a mixture of 75 (65) vol.\% of amorphous carbon and 25 (35) vol.\% of silicate.
To minimise effects from the resonant oscillations of monomers typical of $X$ $\sim$ 1 and from the absence of possibly smaller-sized monomers than applied in computation here, we mainly consider the longward end of the wavelength range we cover (7,000--9,000 \AA). Additionally, the approximated nature of the dust (e.g., non-hierarchical structure without the surface roughness) could reduce the accuracy of the estimation around the minimum polarisation degree and the inversion angle (e.g. \citealt{Kolokolova2018}), and indeed produces an unlikely (i.e. too positive) $P$ distribution in our simulation compared to the ordinary cometary dust.
Hence, given that a general polarimetric phase curve of cometary dust in the NPB (0\degree--22\degree) is roughly symmetrical around the $P_{\rm min}$ at $\alpha_{\rm min}$ $\sim$ 11\degree \citep{Kiselev2015}, we consider modelled data only on the left side of the curve (i.e. $\alpha$ = 1\degree--10\degree) to minimise the impact of our simple properties of the modelled dust. In order to see a general trend, we take a mean of the modelled $P_{\rm r}$ values over the considered $\alpha$ range and mark it as a filled circle. Their error bars denote the standard deviations of the $P_{\rm r}$ values.

Dust with a porosity of 60 \% shows the best compatibility among the three modelled dust particles to the observed neutral $PC$ (Fig. \ref{Fig05}) and the $P_{\rm r}$ values at the phase angles in 2016 February and March (Fig. \ref{Fig06}), within the error bars. Overall, the more transparent the dust, the greater the difference in $P_{\rm r}$ according to the change of porosity. For the low-porosity case (60 \%), the $PC$ is not significantly different between the two modelled cases within the error bars, but the depth of the NPB for the transparent dust (grey symbols in Fig. \ref{Fig10}) is shallower than the 67P observations. For the fluffy case (95 \%), the transparent dust shows a more negative $P_{\rm r}$ and steeper blue $PC$ than the absorbing dust (black symbols in Fig. \ref{Fig10}). The transparent dust shows either too shallow (for the 60 \% case with the average $P_{\rm r}$ of $-$0.2 \%) or too deep (for the 95 \% case with the average $P_{\rm r}$ of $-$3.4 \%) depth in the NPB compared to the observed values of 67P dust. The transparent dust of a 80 \% porosity (grey dashed lines in panel d) also showed a local deep of $P_{\rm r}$ at 8,000 \AA\ and large $P_{\rm r}$ difference between 7,000 \AA\ and 9,000 \AA, both of which are incompatible to our observed $PC$ (Figs. 1--3). 
For these reasons, we will only consider the latter (i.e. a mixture of 75 vol.\% of amorphous carbon and 25 vol.\% of Mg-rich silicate) in the following discussion. 
$P_{\rm r}$ of dust with a porosity of 80 \% appears to extend toward a bit more negative side in the NPB but still could explain the observed $PC$ reasonably well. The 95 \% dust porosity explains neither the neutral $PC$ nor the $P_{\rm r}$ values of the observed results. As such, we conclude that the dust particles dominating around the end of the summer are consistent with dust agglomerate with the porosity of 60--80 \% but with a better match on the 60 \%-end side. In the following, we use 60 \% as a representative porosity value of the dust particles, showing non-fragmentation behaviour \citep{Hornung2016}. The retrieved microporosity of the dust particles is lower than the macroscopic porosity of the 67P nucleus of 75--85 \% estimated by CONSERT  \citep{Kofman2015,Kofman2020}, 65--80 \% from the Rosetta/Radio Science Investigation (RSI) \citep{Patzold2019} and 70--75 \% from the Optical, Spectroscopic, and Infrared Remote Imaging System (OSIRIS) pre-perihelion observations \citep{Jorda2016}. The microporosity is also lower than the macroscopic surface porosity of 87 \% derived by the Hapke model \citep{Fornasier2015}.

The mean density of dust particles with the porosity of 60 \% is then 
\begin{equation}
\bar{\rho} = \rho_{\rm c}f_{\rm c} + \rho_{\rm Si}f_{\rm Si}~,
\label{eq:eq15}
\end{equation}
\noindent where
\begin{equation}
f_{\rm c} + f_{\rm Si} + f_{\rm void} = 1~.
\label{eq:eq16}
\end{equation}
$\rho_{\rm X}$ and $f_{\rm X}$ are the density and fractional volumes occupied by the component X, respectively, where X equals `c' for amorphous carbon, `Si' for Mg-rich silicates, and `void' for the void between the constituent particles we constrained. By substituting $\rho$ = 3,360 kg m$^{\rm -3}$ for the used silicates\footnote{obtained by \citet{Emery2002} using the compositional information from \citet{Dorschner1995}}, 1,435 kg m$^{\rm -3}$ for the amorphous carbon \citep{Jager1998}, $f_{\rm c}$ = 0.3, $f_{\rm Si}$ = 0.1, and $f_{\rm void}$ = 0.6 into the equations, we obtain the mean density of the agglomerate as $\bar{\rho}$ $\sim$ 770 kg m$^{\rm -3}$. Although the exact value could be different for dust constructed more sophisticatedly, the estimated density here is undoubtedly higher than the bulk density of the 67P nucleus (535 $\pm$ 35 kg m$^{\rm -3}$; \citealt{Jorda2016}) but well consistent with that of compact particles detected by GIADA (785$^{\rm +520}_{\rm -115}$ kg m$^{\rm -3}$; \citealt{Fulle2017}).

This type of dust particles is expected to be ubiquitous on the surfaces of short-period comets \citep{Hadamcik2009}, which would be the outcome of the perennial sintering process under solar radiation. 
The predominance of such processed compact particles then explains the absence of 10 $\mu$m silicate emission features of 67P dust coma \citep{Hanner1985} based on the positive correlation between the dust porosity and the silicate emission excess \citep{Kwon2021}. In the context of dust evolution, the behaviours of 67P dust in the NPB present in this study would probably showcase the ongoing processing of the 67P surface relative to its last apparitions and other dynamic groups of comets.

\subsection{Constraints on the amount of water ice in and the refractory-to-ice ratio of the compact dust particles \label{sec:dis2}}

Although the dust activity level of 67P in early 2016 weakened significantly compared to its apex around the perihelion, the comet was still located inside the water frost line ($r_{\rm H}$ = 2.7 au; \citealt{Jewitt2007}) that the activity was discernible enough. The dust activity of the comet in this season was most likely driven by sublimation of water-ice \citep{Hansen2016}. The heliocentric distance ($r_{\rm H}$ $\sim$ 2.5 au) at which we observed 67P is too large for organic materials to sublimate (usually at $r_{\rm H}$ $\lesssim$ 0.7 au; \citealt{Jones2018}) but too small for bombarding solar wind particles to evaporate ice particles (significant at $r_{\rm H}$ $>$ 5 au; \citealt{Mukai1986}), and for supervolatile ices (e.g. CO$_{\rm 2}$ and CO) to lift off the constrained dust particles into the coma \citep{Jewitt2015}. \citet{Frattin2017} suggest from the OSIRIS observations that the post-perihelion inner coma dust of 67P consists of three primary components --- organics, silicates, and water ice.  A ground-based observation with the VLT/MUSE also corroborates the dust activity mainly driven by water-ice sublimation in early 2016 via clustering of the [OI] emission line at 6,300 \AA\ around the nucleus \citep{Opitom2020}.

The dust colour is the ratio of the intensity measured at two different wavelengths, thus able to trace the attributes of the dust particles. Although the Rosetta spacecraft was not equipped with a polarimeter, the dust colour measured at different positions in the coma by onboard instruments allows to investigate the radial evolution of the dust properties in the coma of 67P. To this end, we utilise the Visible and Infrared Thermal Imaging Spectrometer (VIRTIS) spectroscopic data \citep{Filacchione2020}, and OSIRIS multi-band imaging data \citep{Frattin2017} measured at 1.0-2.5 km and 80--437 km in cometocentric distance, respectively. Strictly speaking, our colour data is the integration of all signals within the 1,000 km aperture, not the local signal at 1,000 km. We could nonetheless assume that our data well approximates the dust environment around 1,000 km in radial distance, based on (i) both the polarimetric and colour data stable against the effects induced by the near-nucleus outburst event in 2016 Feb \citep{Grun2016}; and (ii) apparently no changes over the two months indicating the sampling of the steady-state coma dust where the ejection information would have vanished out of the aperture.

We digitised the data points in panels (b) and (d) of Fig. 1 in \citet{Filacchione2020} and multiply 10 to match the unit. The data from Fig. 7 in \citet{Frattin2017} were obtained around UT 2016 January, yet would not be problematic under the almost identical dust colour between January and March (Fig. \ref{Fig09}a). The exact start and end wavelength positions of the blue and red sides are different between the observations: \citet{Filacchione2020} set 4,000--5,000 \AA\ for the blue side and 5,000--8,000 \AA\ for the red side,  \citet{Frattin2017} set 4,800--6,490 \AA\ for the blue side and 6,490--8,820 \AA\ for the red side, and we set 4,000--5,500 \AA\ and 5,600--7,900 \AA\ for the blue and red sides, respectively. However, the spectral region where water ice shows its highest transparency (at $\sim$4,000--5,000 \AA; \citealt{Mukai1986,Warren2019}) is always covered by the blue sides of the observations, such that we could reliably monitor any compositional variation induced by the water-ice sublimation. The change in the absorptivity of the mixture of the other two major materials (silicates and carbonaceous components) is continuous over 4,000--9,000 \AA\ to a large extent \citep{Dorschner1995,Kolokolova1997b}. The differences in $\alpha$ between the observations (on average $\sim$90\degree, $\sim$90\degree, and $\sim$17\degree\ from inner to outer coma radius) would change the exact value of reddening. However, we expect that a weak ($\lesssim$1 \%) dependence of the dust colour on the phase angle  \citep{Jewitt1986,Gustafson1999,Kimura2003,Kolokolova2004,Bertini2017} keeps the observed reddening trend valid.

\begin{figure}[!htb]
\centering
\includegraphics[width=9cm]{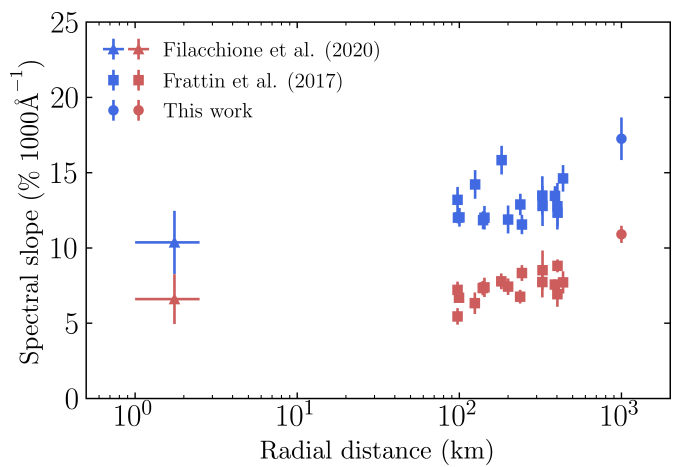}
\caption{The radial evolution of the spectral slope of the dust particles of 67P in \% (1,000 \AA)$^{\rm -1}$. Circles indicate the average value of our three-epoch colour data (Fig. \ref{Fig09}a). Blue and red symbols indicate spectral slopes measured at wavelengths shortward and longward of the inflection point. Horizontal error bars of triangles span the cometocentric distance (1.0--2.5 km) VIRTIS covered, while their vertical error bars denote the error of the weighted mean of observations (i.e. the standard deviation divided by the square root of the number of the data points) between UT 2016 January and early March. }
\label{Fig11}
\end{figure}

Fig. \ref{Fig11} shows a synthesis of the spectral slopes in \% (1,000 \AA)$^{\rm -1}$ measured at different cometocentric distances. The reddening of the dust as it moves outward the nucleus is clearly shown. Provided that colour changes result from the changes in dust particles' compositional and/or physical properties  \citep{Kolokolova2004}, we could first rule out the latter effect since the Rosetta observations confirm the absence of any systematic shift of dust agglomerates into the smaller size distribution in the process of dust moving out of the nucleus at least in the $<$500 km distance range \citep{Fulle2016,Frattin2021}. Then, in terms of change in the dust composition, the observed trend in colour could support the sublimation of water-ice particles in the coma of 67P inferred from the Rosetta observations in early 2016. The different degree of reddening in the blue and red sides is also consistent with this scenario. All blue sides in Fig. \ref{Fig11} cover the 4,000--5,000 \AA\ region (where the water ice has a minimum in the imaginary index of refraction); thus, it would be natural that the sublimation impact would be more significant therein than on the red side.

We performed an order-of-magnitude estimation to conjecture the possible amount of water-ice embedded in the dust particles and therefrom their refractory-to-ice ratio. Most of the particles move radially from the nucleus \citep{Frattin2021}. For simplicity in the calculation, we assume that (i) the sublimation starts on ejection (0 km in altitude) and ends at 1,000 km, roughly consistent with the observed central clustering of the intensity of the [OI] line \citep{Opitom2020}; (ii) water ice exists as pure one, which might be a proper treatment based on the Rosetta observations (e.g. \citealt{Filacchione2016}); and (iii) pure water-ice and refractory dust coexist not in thermal contact \citep{Yang2009} such that the ice keeps its low temperature during sublimation. Given the velocity of 100 $\mu$m dust of $\sim$order of 10 m s$^{\rm -1}$ \citep{Fulle2010,Grun2016}, it takes approximately 10$^{\rm 5}$ s to reach the endpoint. Any size of 1--1,000 $\mu$m of pure ice might survive for the time at $r_{\rm H}$ $\sim$ 2.5 au \citep{Mukai1986}. The rate of size decrease of the ice by sublimation can be expressed as
\begin{equation}
\bigg|\frac{\ud a}{\ud t}\bigg| = \bigg(\frac{\mu m_{\rm p}}{\rho}\bigg)~p_{\rm v}(T) (2\pi \mu m_{\rm p} k_{\rm B} T)^{\rm -1/2}
\label{eq:eq17}
\end{equation}
\citep{Mukai1981}, where $a$ is the radius of the dust particle, $t$ is the time (10$^{\rm 5}$ s), $\mu$ is the mean molecular weight of water ice, $m_{\rm p}$ is its atomic mass unit, $\rho$ is the mass density of water-ice (1,000 kg m$^{\rm -3}$ is used), $k_{\rm B}$ is the Boltzmann's constant, and $T$ is the water-ice temperature at the given heliocentric distance ($\sim$120 K for pure ice; \citealt{Mukai1986}). $p_{\rm v}$(T) means the saturated vapour pressure of the water ice at $T$ for which we adopted the equation of \citet{Lamy1974} valid at low temperature
\begin{equation}
\begin{aligned}
\log p_{\rm v}(T) = {} & -\frac{2461}{T} + 3.857 \log T + 3.41 \times 10^{\rm -3} T \\
& + 4.875 \times 10^{\rm -8}T^{\rm 2} + 4.332~.
\label{eq:eq18}
\end{aligned}
\end{equation}
\noindent Substituting the values for pure water ice into Eqs. \ref{eq:eq17} and \ref{eq:eq18}, the size of water ice for 10$^{\rm 5}$ s reduces by $\sim$40 $\mu$m in linear dimension. Given that even the non-fragmenting dust agglomerates exhibit their inner structure quite well \citep{Hornung2016}, we assume that the water ice resides top 40 $\mu$m crust of the 100-$\mu$m-sized spherical dust agglomerate (the dust considered in Sect. \ref{sec:dis1}). This results in the radius of an initially ejected dust particle (before sublimation) to be $\sim$140 $\mu$m{\footnote{If the ice-bearing constituents have an intimate mixture of ice and refractories in noticeable thermal contact, unlike assumptions herein, then the resultant size of the dust particles needs to be centimetre-sized or larger \citep{Mukai1986,Markkanen2020}.}}. To make such dust plausible albedo-wise, by taking into account the volume fraction of ice abundance of the bright spots on the 67P surface on average of 4 vol.\% (1--7.2 vol.\%; \citealt{Barucci2016}), we further assume that 4 \% of the volume of the top crust consists of the ice particles. As a result, we obtain the volume fraction of water ice in the initially ejected dust as $\sim$6 \% and their refractory-to-ice ratio of $\sim$14. The dust density is then 
\begin{equation}
\rho_{\rm init} = \rho_{\rm c}f_{\rm c} + \rho_{\rm Si}f_{\rm Si} + \rho_{\rm ice}f_{\rm ice}
~,
\label{eq:eq19}
\end{equation}
\noindent where  
\begin{equation}
f_{\rm c} + f_{\rm Si} + f_{\rm ice} + f_{\rm void, init} = 1 
~.
\label{eq:eq19}
\end{equation}
Here $f_{\rm void, init}$ denotes the initial void volume fraction (before sublimation), obtained by subtracting the ice fraction ($f_{\rm ice}$) from the constrained porosity (0.6) in Sect. \ref{sec:dis1}. $f_{\rm c}$ = 0.3 and $f_{\rm Si}$ = 0.1 are used as in the previous section. The resultant $\rho_{\rm init}$ is $\sim$830 kg m$^{\rm -3}$. Since the volume of the dust particles is mostly occupied by refractory materials, the change of mechanical strength after the sublimation would be negligible \citep{Haack2020}. Assuming that the tensile strengths of water-ice and organic materials in the dust are not significantly different from that of silicates \citep{Gundlach2018,Bischoff2020}, from Eq. 5 in \citet{Meisner2012}, our filling factor of the dust upon ejection (1 $-$ $f_{\rm void, init}$) corresponds to its tensile strength of $\sim$8.2 kPa. This estimated tensile strength is much higher than the result of the dust showing catastrophic fragmentation patterns but well consistent with the one of dust showing no or slight fragmentations in the collection of the COSIMA experiment \citep{Hornung2016,Langevin2016}.
\\

\section{Summary \label{sec:sum}}

We present new VLT spectropolarimetric observations of comet 67P/Churyumov-Gerasimenko obtained at the end of its intense Southern summer. These datasets give us simultaneous polarimetric and colorimetric views at optical wavelengths. The main results of the analysis are as follows.

\begin{enumerate}

\item  The $P_{\rm r}$ of 67P dust is positive at $\alpha$ = 26\fdg24 on UT 2016 Jan 12, and negative at $\alpha$ = 18\fdg47 and 5\fdg49 on UT 2016 Feb 04 and UT 2016 Mar 04, respectively, as expected from the observations of other cometary dust particles. Their wavelength dependences $P_{\rm r}(\lambda)$ over 4,000--9,000 \AA\ are  (0.14 $\pm$ 0.08) \% (1,000 \AA)$^{\rm -1}$, ($-$0.03 $\pm$ 0.07) \% (1,000 \AA)$^{\rm -1}$, and (0.03 $\pm$ 0.12) \% (1,000 \AA)$^{\rm -1}$ for the three dates, which are much flatter than those of the dust of other comets, especially long-period comets and 2I/Borisov at similar observing geometries.\\

\item The trigonometric function describing the dependence of 67P dust's $P_{\rm r}$ on the phase angle $P_{\rm r}(\alpha)$ can be parameterised by $b$ = 52.32 $\pm$ 20.87 \%, $c_{\rm 1}$ = 1.01 $\pm$ 0.24, $c_{\rm 2}$ = 9.99 $\pm$ 5.82, and $\alpha_{\rm 0}$ = 21\fdg71 $\pm$ 0\fdg51. The limited $\alpha$ coverage ($<$40\degree) and the small number of the available 67P data yields large errors of the fitting parameter $b$, which is relevant to the amplitude of the phase curve. The post-perihelion data of 67P dust appear more strongly polarised than pre-perihelion at $\alpha$ $>$ 22\degree, but the number of pre-perihelion data sets is insufficient to make any firm conclusions. The depth of the NPB of 67P dust is slightly shallower than those of the long- and non-period comets, but comparable to other short-period comets.\\

\item The absolute value of the $P_{\rm r}$ at the given $\alpha$ in the NPB appears to decrease (i.e. a decrease in the depth of the NPB) between apparitions. The weighted mean of the $P_{\rm r}$ departure from the average trend is ($-$0.30 $\pm$ 0.10) \%, (0.25 $\pm$ 0.13) \%, and (0.46 $\pm$ 0.15) \% for the 1983, 2009, and 2015 apparitions, respectively.\\

\item The 67P's dust colour remains nearly constant over the two months in early 2016. It shows red colour, with an inflection point around 5500 \AA. The normalised spectral slopes of the shortward (blue side; 4,000--5,000 \AA) and longward (red side; 5,600--7,900 \AA) of the point are (17.3 $\pm$ 1.4) \% (1,000 \AA)$^{\rm -1}$ and (10.9 $\pm$ 0.6) \% (1,000 \AA)$^{\rm -1}$, respectively. 67P during this term displays a slightly redder but overall consistent dust colour with the average of active Jupiter-Family comets.\\

\item Our observations of 67P in early 2016 are compatible with the light scattering behaviour of the non-fragmenting dust agglomerates retrieved from the Rosetta observations. To estimate the plausible porosity of the dust, we construct a simple light scattering model, setting a spherical dust aggregate particle of 100 $\mu$m in size. The dust consists of spherical monomers following a power-law size distribution with the differential index of $-$3 and the minimum and maximum sizes of 0.05 $\mu$m and 0.25 $\mu$m, respectively. Each monomer represents one type of composition, either amorphous carbon or silicate, constituting 75 vol.\% and 25 vol.\% in refractory materials, respectively. The dust porosity of $\sim$60 \% shows the best match with our polarimetric results. We also test a dust mixture of 65 vol.\% of amorphous carbon and 35 vol.\% of silicate, but its polarimetric properties are found to be inconsistent with the observed $PC$ and $P_{\rm r}$ of 67P dust. The resultant density is $\sim$770 kg m$^{\rm -3}$, higher than the bulk density of the 67P nucleus but consistent with the density of compact particles detected by the GIADA experiment.\\

\item Compiling the colour data obtained from the Rosetta (VIRTIS and OSIRIS) and our ground-based observations in early 2016, we find the signature of water-ice sublimation and the ensuing reddening of the dust colour as the radial distance increases. From an order-of-magnitude estimation, the volume ratio of water-ice embedded in the dust particles before sublimation is expected to be $\sim$6 \%, yielding the refractory-to-ice volume ratio of the dust as $\sim$14. Consequently, its density and tensile strength would be $\sim$830 kg m$^{\rm -3}$ and $\sim$8.2 kPa, respectively. The retrieved tensile strength is consistent with the one derived from the dust particles showing no or slight fragmentation patterns in the COSIMA collection.

\end{enumerate}

The observations presented here suggest the great potential of polarimetry as a tool to constrain physical properties of cometary dust and examine any differences therein between less-altered non- and long-period comets and relatively sintered short-period comets. We anticipate that this observing mode would help characterise dust profiles of a large number of comets in the years to come, especially for which only ground-based observations are available.
\\

\begin{acknowledgements}

We thank Gianrico Filacchione and Elisa Frattin for providing the Rosetta VIRTIS and OSIRIS points in Fig. \ref{Fig11}. YGK gratefully acknowledges the support of the Alexander von Humboldt Foundation. JA  acknowledges funding by the Volkswagen Foundation. JA. and JM. acknowledge funding from the European Union’s Horizon 2020 research and innovation programme under grant agreement No 757390 CAstRA. ACLR acknowledges support from the French Space Agency (CNES), in relation with Rosetta and Comet Interceptor missions. CS acknowledges funding from the UK STFC, grants ST/L004569/1 and ST/V000586/1. This study is based on observations collected at the European Southern Observatory under ESO programme 096.C-0821.

\end{acknowledgements}


\begin{appendix}
\section{The extracted instrumental spectra of 67P with different aperture sizes \label{sec:app}}

\begin{figure}[!htb]
\centering
\includegraphics[width=9cm]{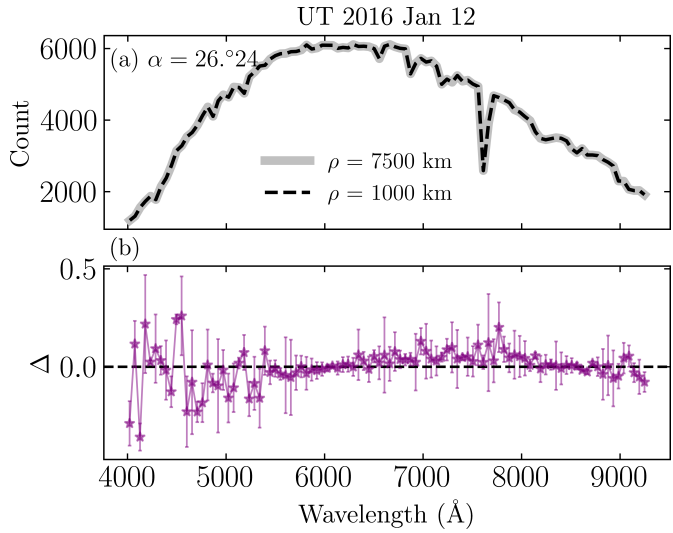}
\caption{Counts of 67P dust extracted with the aperture sizes corresponding to 1,000 km (black) and 7,500 km (grey) in the cometocentric distance (a) and their differences (b) as a function of wavelength on UT 2016 Jan 12. $\alpha$ indicates the phase angle on the epoch. In panel (a), black and grey lines overlap that they are indistinguishable from visual inspection.}
\label{Fig12}
\vskip-1ex
\end{figure}

\begin{figure}[!htb]
\centering
\includegraphics[width=9cm]{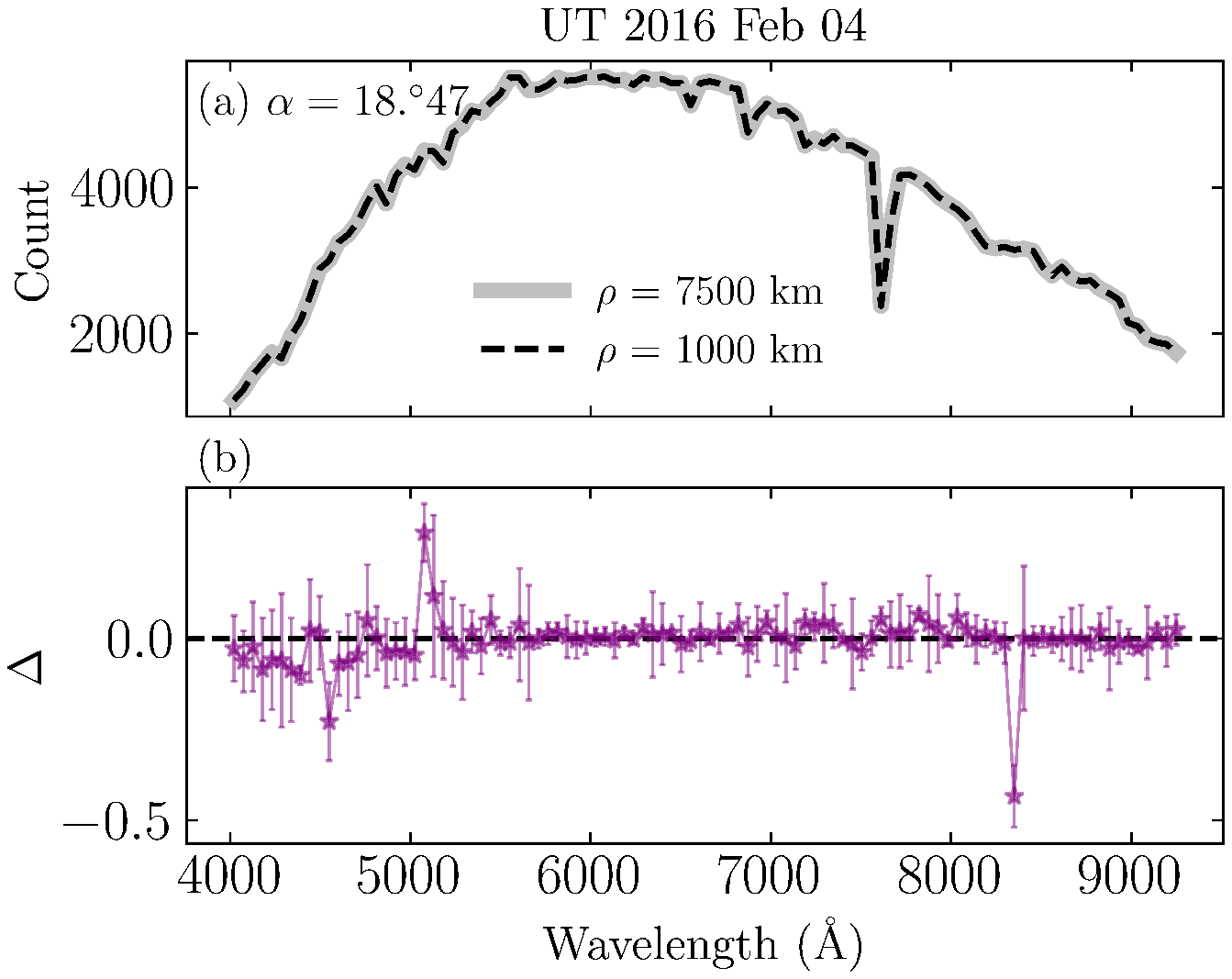}
\caption{Same as Figure \ref{Fig12} but on UT 2016 Feb 04. Black and grey lines overlap each other.}
\label{Fig13}
\vskip-1ex
\end{figure}

\begin{figure}[!htb]
\centering
\includegraphics[width=9cm]{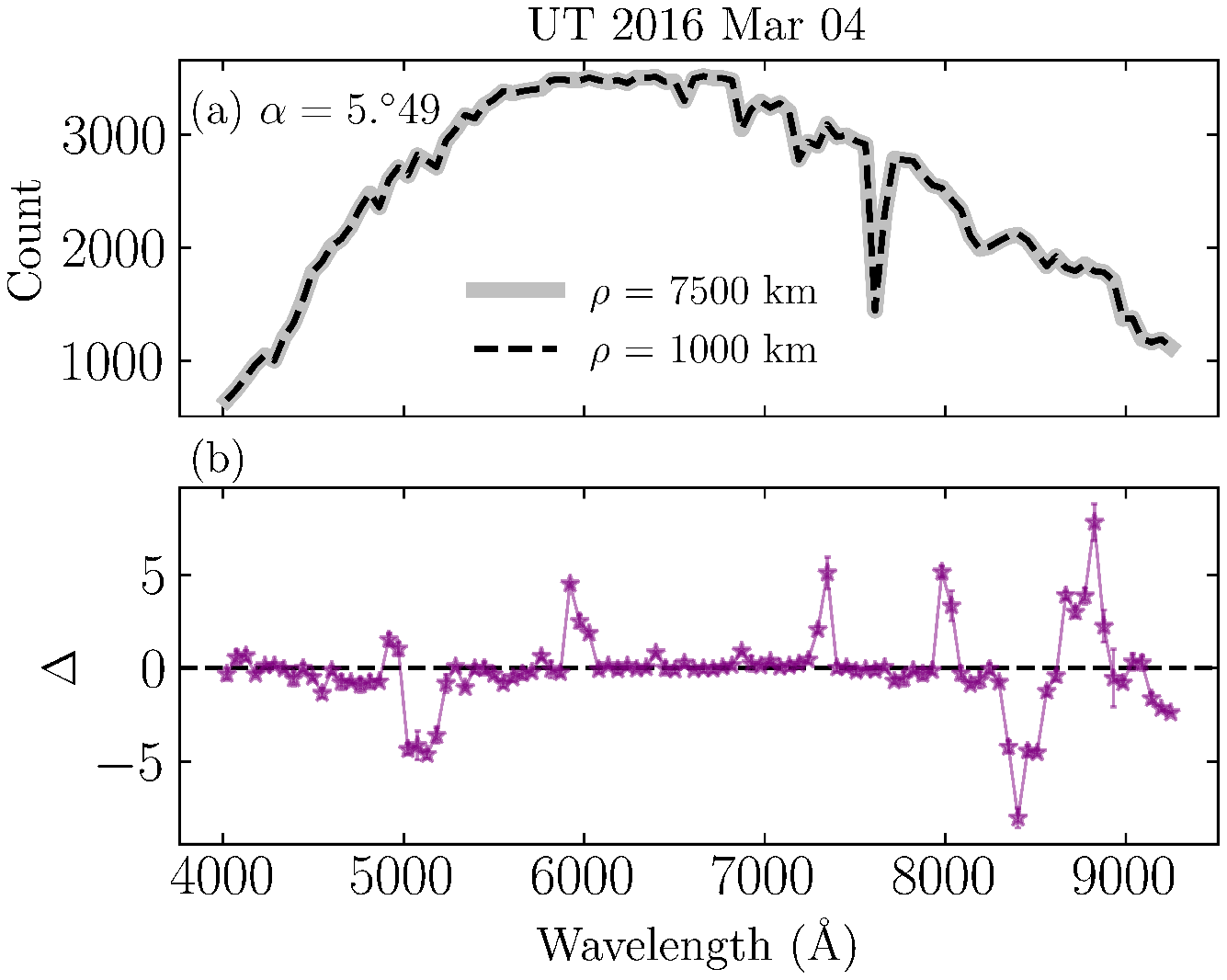}
\caption{Same as Figure \ref{Fig12} but on UT 2016 Mar 04. Black and grey lines overlap each other.}
\label{Fig14}
\vskip-1ex
\end{figure}

The extracted sixteen light elements on each epoch are summed to retrieve Stokes $I$ using Eq. \ref{eq:eq3}. In six different aperture sizes applied, each light component and synthesised Stokes $I$ show no radial variations both in the count and spectral shape throughout the observations. Figs. \ref{Fig12}--\ref{Fig14} present $I$ measured at 1,000 km and 7,500 km (panel (a)) and their differences (panel (b)) for the January, February, and March observations, respectively. Black (1,000 km aperture size) and grey (7,500 km aperture size) lines are overlapped, making their differences ($\Delta$) almost zero. The larger the aperture sizes, the more significant error becomes due to the enhanced background noise. Thus, we decided to make use of the spectrum extracted with the 1,000-km-sized aperture for the observations at all observing epochs. 


\end{appendix}


\begin{thebibliography}{}

\bibitem[Agarwal et al.(2016)]{Agarwal2016} Agarwal, J., A'Hearn, M. F., Vincent, J.-B., et al. 2016, \mnras, 462, S78 
\bibitem[A'Hearn(1984)]{A'Hearn1984} A'Hearn, M. A. 1984, \aj, 89, 579
\bibitem[Bagnulo et al.(2009)]{Bagnulo2009} Bagnulo, S., Landolfi, M., Landstreet, J. D., et al. 2009, \pasp, 121, 993
\bibitem[Bagnulo et al.(2017)]{Bagnulo2017} Bagnulo, S., Cox, N. L. J., Cikota, A., et al. 2017, A\&A, 608, 146
\bibitem[Bagnulo et al.(2021)]{Bagnulo2021} Bagnulo, S., Cellino, A., Kolokolova, L., et al. 2021, Nat. Commun., 12, 1797
\bibitem[Bardyn et al.(2017)]{Bardyn2017} Bardyn, A., Baklouti, D., Cottin, H., et al. 2017, \mnras, 469, S712
\bibitem[Barucci et al.(2016)]{Barucci2016} Barucci, M. A., Filacchione, G., Fornasier, S., et al. 2016, A\&A, 595, A102
\bibitem[Bentley et al.(2016)]{Bentley2016} Bentley, M. S., Schmied, R., Mannel, T., et al. 2016, \nat, 537, 73
\bibitem[Bertini et al.(2017)]{Bertini2017} Bertini, I., La Forgia, F., Tubiana, C., et al. 2017, \mnras, 469, S404
\bibitem[Bertini et al.(2019)]{Bertini2019} Bertini, I., La Forgia, F., Fulle, M., et al. 2019, \mnras, 482, 2924
\bibitem[Biele et al.(2015)]{Biele2015} Biele, J., Ulamec, S., Maibaum, M., et al. 2015, Science, 349, aaa9816 
\bibitem[Bischoff et al.(2020)]{Bischoff2020} Bischoff, D., Kreuzig, C., Haack, D., Gundlach, B., \& Blum, J. 2020, \mnras, 497, 2517
\bibitem[Boehnhardt et al.(2016)]{Boehnhardt2016} Bohenhardt, H., Riffeser, A., Kluge, M., et al. 2016, \mnras, 462, 376
\bibitem[Bohren \& Huffman(1983)]{Bohren1983} Bohren, C., \& Huffman, D. 1983, Absorption and Scattering of Light by Small Particles (New York: Wiley Sons)
\bibitem[Chernova et al.(1993)]{Chernova1993} Chernova, G. P., Kiselev, N. N., \& Jockers, K. 1993, \icarus, 103, 144
\bibitem[Cikota et al.(2017)]{Cikota2017} Cikota, A., Patat, F., Cikota, S., \& Faran, T. 2017, \mnras, 464, 4146
\bibitem[Datson et al.(2015)]{Datson2015} Datson, J., Flynn, C., \& Portinari, L. 2015, A\&A, 574, A124
\bibitem[Della Corte et al.(2015)]{DellaCorte2015} Della Corte, V., Rotundi, A., Fulle, M., et al. 2015, A\&A, 583, A13
\bibitem[Dorschner et al.(1995)]{Dorschner1995} Dorschner, J., Begemann, B., Henning, T., J{\"a}ger, C., \& Mutschke, H. 1995, A\&A, 300, 503
\bibitem[Emery(2002)]{Emery2002} Emery, J. 2002, PhD Thesis, the University of Arizona
\bibitem[Filacchione et al.(2016)]{Filacchione2016} Filacchione, G., De Sanctis, M. C., Capaccioni, F., et al. \nat, 529, 368
\bibitem[Filacchione et al.(2020)]{Filacchione2020} Filacchione, G., Capaccioni, F., Ciarniello, M., et al. \nat, 578, 49
\bibitem[Fornasier et al.(2015)]{Fornasier2015} Fornasier, S., Hasselmann, P., Barucci, M. A., et al. 2015, A\&A, 583, A30
\bibitem[Fulle et al.(2010)]{Fulle2010} Fulle, M., Colangeli, L., Agarwal, J., et al. 2010, A\&A, 522, A63
\bibitem[Fulle et al.(2016)]{Fulle2016} Fulle, M., Altobelli, N., Buratti, B., et al. 2016, \mnras, 462, 2
\bibitem[Fulle et al.(2017)]{Fulle2017} Fulle, M., Della Corte, V., Rotundi, A., et al. 2017, \mnras, 469, S45
\bibitem[Fulle et al.(2020)]{Fulle2020} Fulle, M., Blum, J., Rotundi, A., et al. 2020, \mnras, 493, 4039
\bibitem[Frattin et al.(2017)]{Frattin2017} Frattin, E., Cremonese, G., Simioni, E., et al. 2017, \mnras, 469, S195
\bibitem[Frattin et al.(2021)]{Frattin2021} Frattin, E., Bertini, I., Ivanovski, S. L., et al. 2021, \mnras, 504, 4687
\bibitem[Gr{\"u}n et al.(2016)]{Grun2016} Gr{\"u}n, E., Agarwal, J., Altobelli, N., et al. 2016, \mnras, 462, S220
\bibitem[Gundlach et al.(2018)]{Gundlach2018} Gundlach, B., Schmidt, K. P., Kreuzig, C., et al. 2018, \mnras, 479, 1273
\bibitem[Gundlach et al.(2020)]{Gundlach2020} Gundlach, B., Fulle, M., \& Blum, J. 2020, \mnras, 493, 3690
\bibitem[Gustafson \& Kolokolova(1999)]{Gustafson1999} Gustafson, B. \AA. S., \& Kolokolova, L. 1999, \jgr, 104, 31711
\bibitem[G{\"u}ttler et al.(2019)]{Guttler2019} G{\"u}ttler, C., Mannel, T., Rotundi, A., et al. 2019, A\&A, 24, 1
\bibitem[Haack et al.(2020)]{Haack2020} Haack, D., Otto, K., Gundlach, B., et al. 2020, A\&A, 642, A218
\bibitem[Hadamcik \& Levasseur-Regourd(2003a)]{Hadamcik2003a} Hadamcik, E., \& Levasseur-Regourd, A. C. 2003a, \jqsrt, 79, 661
\bibitem[Hadamcik \& Levasseur-Regourd(2003b)]{Hadamcik2003} Hadamcik, E., \& Levasseur-Regourd, A. C. 2003b, A\&A, 403, 757
\bibitem[Hadamcik \& Levasseur-Regourd(2009)]{Hadamcik2009} Hadamick, E., \& Levasseur-Regourd, A. C. 2009, P\&SS, 57, 1118
\bibitem[Hadamcik et al.(2010)]{Hadamcik2010} Hadamcik, E., Sen, A. K., Levasseur-Regourd, A. C., Gupta, R., \& Lasue, J. 2010, A\&A, 517, A86
\bibitem[Hadamcik et al.(2016)]{Hadamcik2016} Hadamcik, E., Levasseur-Regourd, A. C., Hines, D. C., et al. 2016, \mnras, 462, 507
\bibitem[Hanner et al.(1985)]{Hanner1985} Hanner, M. S., Tedesco, E., Tokunaga, A. T., et al. 1985, \icarus, 64, 11
\bibitem[Hansen et al.(2016)]{Hansen2016} Hansen, K. C., Altwegg, K., Berthelier, J.-J., et al. 2016, \mnras, 462, S491
\bibitem[Hornung et al.(2016)]{Hornung2016} Hornung, K., Merouane, S., Hilchenbach, M., et al. 2016, P\&SS, 133, 63
\bibitem[Ivanova et al.(2017)]{Ivanova2017} Ivanova, O., Rosenbush, V., Kiselev, N., Afanasiev, V., \& Korsun, P. 2017, \mnras, 469, 386
\bibitem[J{\"a}ger et al.(1998)]{Jager1998} J{\"a}ger, C., Mutschke, H., \& Henning, T. 1998, A\&A, 332, 291
\bibitem[Jewitt \& Meech(1986)]{Jewitt1986} Jewitt, D., \& Meech, K. J. 1986, \aj, 310, 937
\bibitem[Jewitt et al.(2007)]{Jewitt2007} Jewitt, D., Chizmadia, L., Grimm, R., \& Prialnik, D. 2007, in Reipurth, V. B., Jewitt, D., Keil, K., eds, Protostars and Planets V. Univ. Arizona Press, Tucson, 863
\bibitem[Jewitt(2015)]{Jewitt2015} Jewitt, D. 2015, \aj, 150, 201
\bibitem[Jones et al.(2018)]{Jones2018} Jones, G. H., Knight, M. M., Battams, K., et al. 2018, \ssr, 214, 20
\bibitem[Jorda et al.(2016)]{Jorda2016} Jorda, L., Gaskell, R., Capanna, C., et al. 2016, \icarus, 277, 257
\bibitem[Kikuchi(2006)]{Kikuchi2006} Kikuchi, S. 2006, \jqsrt, 100, 179
\bibitem[Kimura et al.(2003)]{Kimura2003} Kimura, H., Kolokolova, L., \& Mann, I. 2003, A\&A, 407, L5
\bibitem[Kimura et al.(2006)]{Kimura2006} Kimura, H., Kolokolova, L., \& Mann, I. 2006, A\&A, 449, 1243
\bibitem[Kiselev \& Velichko(1996)]{Kiselev1996} Kiselev, N., \& Velichko, F. 1996, EM\&P, 78, 347
\bibitem[Kiselev et al.(2015)]{Kiselev2015} Kiselev, N., Rosenbush, V., Kolokolova, L., \& Levasseur-Regourd, A. C. 2015, in Polarimetry of Stars and Planetary Systems, ed. L. Kolokolova, J. Hough, \& A. C. Levasseur-Regourd (Cambridge: Cambridge Univ. Press), 379
\bibitem[Kiselev et al.(2017)]{Kiselev2017} Kiselev, N., Shubina, E., Velichko, S., Jockers, K., Rosenbush, V., and Kikuchi, S. (eds.), Compilation of Comet Polarimetry from Published and Unpublished Sources, urn:nasa:pds:compil-comet:polarimetry::1.0, NASA Planetary Data System, 2017
\bibitem[Kofman et al.(2015)]{Kofman2015} Kofman, W., Herique, A., Barbin, Y., et al. 2015, Science, 349, 6247
\bibitem[Kofman et al.(2020)]{Kofman2020} Kofman, W., Zine, S., Herique, A., et al. 2020, \mnras, 497, 2616
\bibitem[Kolokolova et al.(1997)]{Kolokolova1997} Kolokolova, L., Jockers, K., Chernova, G., \& Kiselev, N. 1997, \icarus, 126, 351
\bibitem[Kolokolova \& Jockers(1997)]{Kolokolova1997b} Kolokolova, L., \& Jockers, K. 1997, PSS, 45, 1543
\bibitem[Kolokolova et al.(2001)]{Kolokolova2001} Kolokolova, L., Jockers, K., Gustafson, B. \AA, S., \& Lichtenberg, G. 2001, \jqsrt, 106, 10113
\bibitem[Kolokolova et al.(2004)]{Kolokolova2004} Kolokolova, L., Hanner, M. S., Levasseur-Regourd, A. C., \& Gustafson, B. \AA. S. 2004, in Comets II, eds. M. Festou, H. U. Keller, \& H. A. Weaver (Tucson: University of Arizona Press), 577
\bibitem[Kolokolova et al.(2015)]{Kolokolova2015} Kolokolova, L., Das, H. S., Dubovik, O., Lapyonok, T., \& Yang, P. 2015, P\&SS, 116, 30
\bibitem[Kolokolova et al.(2018)]{Kolokolova2018} Kolokolova, L., Nagdimunov, L., \& Mackowski, D. 2018, \jqsrt, 204, 138
\bibitem[Knight et al.(2017)]{Knight2017} Knight, M. M., Snodgrass, C., Vincent, J.-B., et al. 2017, \mnras, 469, 661
\bibitem[Kwon et al.(2018)]{Kwon2018} Kwon, Y. G., Ishiguro, M., Shinnaka, Y., et al. 2018, A\&A, 620, A161
\bibitem[Kwon et al.(2019)]{Kwon2019} Kwon, Y. G., Ishiguro, M., Kwon, J., et al. 2019, A\&A, 629, A121
\bibitem[Kwon et al.(2021)]{Kwon2021} Kwon, Y. G., Kolokolova, L., Agarwal, J., Markkanen, J. 2021, A\&A, 650, L7
\bibitem[Lamy(1974)]{Lamy1974} Lamy, P. L. 1974, A\&A, 35, 197
\bibitem[Langevin et al.(2016)]{Langevin2016} Langevin, Y.. Hilchenbach, M., Ligier, N., et al. 2016, \icarus, 271, 76
\bibitem[Lasue et al.(2009)]{Lasue2009} Lasue, J., Levasseur-Regourd, A. C., Hadamcik, E., \& Alcouffe, G. 2009, \icarus, 199, 129
\bibitem[Levasseur-Regourd et al.(2018)]{Levasseur-Regourd2018} Levasseur-Regourd, A. C., Agarwal, J., Cottin, H., et al. 2018, \ssr, 214, 64
\bibitem[Longobardo et al.(2020)]{Longobardo2020} Longobardo, A., Della Corte, V., Rotundi, A. et al. \mnras, 496, 125
\bibitem[Lumme \& Muinonen(1993)]{Lumme1993} Lumme, K., \& Muinonen, K. 1993, Asteroids, Comets, Meteors, IAU Symp., 160, 194
\bibitem[Meisner et al.(2012)]{Meisner2012} Meisner, T., Wurm, G., \& Teiser, J. 2012, A\&A, 544, A138
\bibitem[Mann et al.(2004)]{Mann2004} Mann, I., Kimura, H., \& Kolokolova, L. 2004, \jqsrt, 89, 291
\bibitem[Mannel et al.(2016)]{Mannel2016} Mannel, T., Bentley, M. S., Schmied, R., et al. 2016, \mnras, 462, S304
\bibitem[Mannel et al.(2019)]{Mannel2019} Mannel, T., Bentley, M. S., Boakes, P. D., et al. 2019, A\&A, 630, A26
\bibitem[Marschall et al.(2020)]{Marschall2020} Marschall, R., Skorov, Y., Zakharov, V., et al. 2020, \ssr, 216, 130
\bibitem[Markkanen \& Agarwal(2020)]{Markkanen2020} Markkanen, J., \& Agarwal, J. 2020, A\&A, 643, A16
\bibitem[Mottola et al.(2015)]{Mottola2015} Mottola, S., Arnold, G., Grotheus, H.-G., et al. 2015, Science, 349, aab0232
\bibitem[Muinonen et al.(2015)]{Muinonen2015} Muinonen, K., Penttila, A., \& Videen, G. 2015, in Polarimetry of Stars and Planetary Systems, ed. L. Kolokolova, J. Hough, \& A. C. Levasseur-Regourd (Cambridge: Cambridge Univ. Press), 114
\bibitem[Muinonen et al.(2018)]{Muinonen2018} Muinonen, K., Markkanen, J., V{\"a}is{\"a}nen, Peltoniemi, J., \& Penttil{\"a}, A. 2018, Opt. Lett, 43, 683
\bibitem[Mukai \& Schwehm(1981)]{Mukai1981} Mukai, T., \& Schwehm, G. 1981, A\&A, 95, 373
\bibitem[Mukai(1986)]{Mukai1986} Mukai, T. 1986, A\&A, 164, 397
\bibitem[Myers \& Nordsiseck(1984)]{Myers1984} Myers, R. V., \& Nordsieck, K. H. 1984, \icarus, 58, 431
\bibitem[Opitom et al.(2017)]{Opitom2017} Opitom, C., Snodgrass, C., Fitzsimmons, A., et al. 2017, \mnras, 469, 222
\bibitem[Opitom et al.(2020)]{Opitom2020} Opitom, C., Guilbert-Lepoutre, A., Besse, S., Yang, B., \& Snodgrass, C. 2020, A\&A, 644, 143
\bibitem[Pajola et al.(2017)]{Pajola2017} Pajola, M., Lucchetti, A., Fulle, M., et al. 2017, \mnras, 469, S636
\bibitem[Patat et al.(2011)]{Patat2011} Patat, F., Moehler, S., O'Brien, K., et al. 2011, A\&A, 527, A91
\bibitem[P{\"a}tzold et al.(2019)]{Patzold2019} P{\"a}tzold, M., Ander, T. P., Hahn, M., et al. 2019, \mnras, 483, 2337
\bibitem[Petrova et al.(2001)]{Petrova2001} Petrova, E. V., Jockers, K., \& Kiselev, N. N. 2001, \ssr, 35, 390
\bibitem[Preusker et al.(2017)]{Preusker2017} Preusker, F., Scholten, F., Matz, K., et al. 2017, A\&A, 607, L1
\bibitem[Raponi et al.(2020)]{Raponi2020} Raponi, A., Ciarniello, M., Capaccioni, F., et al. 2020, Nat. Astron., 5, 500
\bibitem[Renard et al.(1996)]{Renard1996} Renard, J.-B., Hadamcik, E., \& Levasseur-Regourd, A. C. 1996, A\&A, 316, 263
\bibitem[Rosenbush et al.(2017)]{Rosenbush2017} Rosenbush, V., Ivanova, O., Kiselev, N., Kolokolova, L., \& Afanasiev, V. 2017, \mnras, 469, 475
\bibitem[Schleicher(2006)]{Schleicher2006} Schleicher, D. 2006, \icarus, 181, 442
\bibitem[Snodgrass et al.(2016)]{Snodgrass2016} Snodgrass, C., Opitom, C., De Val-Borro, M., et al. 2016, \mnras, 462, 138
\bibitem[Snodgrass et al.(2017)]{Snodgrass2017} Snodgrass, C., A’Hearn, M. F., Aceituno, F., et al. 2017, Phil. Trans. R. Soc. London Ser. A, 375, 20160249
\bibitem[Solontoi et al.(2012)]{Solontoi2012} Solontoi, M., Ivezi{\'c}, {\v Z}., Juri{\'c}, M., et al. 2012, \icarus, 218, 571
\bibitem[Spohn et al.(2015)]{Spohn2015} Spohn, T., Knollenber, J., Ball, A. J., et al. 2015, Science, 349, aab0464
\bibitem[Stinson et al.(2016)]{Stinson2016} Stinson, A., Bagnulo, S., Tozzi, G. P., et al. 2016, A\&A, 594, A110
\bibitem[Storrs et al.(1992)]{Storrs1992} Storrs, A., Cochran, A., \& Barker, E. 1992, \icarus, 98, 163
\bibitem[Thomas et al.(2019)]{Thomas2019} Thomas, N., Ulamec, S., K{\"u}hrt, E., et al. 2017, \ssr, 215, 47
\bibitem[van Dokkum(2001)]{Dokkum2001} van Dokkum, P. G. 2001, \pasp, 113, 1420
\bibitem[V{\"a}is{\"a}nen et al.(2019)]{Vaisanen2019} V{\"a}is{\"a}nen, T., Markkanen, J., Penttil{\"a}, A., \& Muinonen, K. 2019, PloS one, 14, 1
\bibitem[Warren(2019)]{Warren2019} Warren, S. 2019, Phil. Trans. R. Soc. A, 377, 20180161
\bibitem[Yang et al.(2009)]{Yang2009} Yang, B., Jewitt, D., \& Bus, S. J. 2009, \aj, 137, 4538

\end{thebibliography}
\end{document}